\documentclass[prc,aps,superscriptaddress,nofootinbib,showkeys,showpacs,twocolumn]{revtex4-2}
\usepackage{epsfig}
\usepackage{graphicx}
\graphicspath{ {./Pictures/} }
\usepackage{amssymb}
\usepackage{color}
\usepackage{upgreek}
\usepackage{mathtools}
\usepackage{scalerel}
\usepackage{bm}
\usepackage[dvipsnames]{xcolor}
\usepackage{dsfont}
\usepackage{braket}
\usepackage{tikz}
\usetikzlibrary{quantikz2}
\usepackage{array}
\usepackage{comment}

\begin{document}

%\title{Calculation of Green's functions using quantum computers for small superfluid systems}
\title{Quantum simulations of Green's functions for small superfluid systems}

\author{Samuel Aychet-Claisse} \email{samuel.aychet-claisse@universite-paris-saclay.fr}
\affiliation{Universit\'e Paris-Saclay, CNRS/IN2P3, IJCLab, 91405 Orsay, France}
\affiliation{IRFU, CEA, Universit\'e Paris-Saclay, 91191 Gif-sur-Yvette, France}

\author{Denis Lacroix } \email{lacroix@ijclab.in2p3.fr}
\affiliation{Universit\'e Paris-Saclay, CNRS/IN2P3, IJCLab, 91405 Orsay, France}

\author{Vittorio Som\`a}
\email{vittorio.soma@cea.fr}
\affiliation{IRFU, CEA, Universit\'e Paris-Saclay, 91191 Gif-sur-Yvette, France}

\author{Jing Zhang} \email{jing.zhang@ijclab.in2p3.fr}
\affiliation{Universit\'e Paris-Saclay, CNRS/IN2P3, IJCLab, 91405 Orsay, France}

\date{\today}
\begin{abstract}
An end-to-end strategy for hybrid quantum-classical computations of Green's functions in many-body systems is presented and applied to the pairing model.
The scheme makes explicit use of the spectral representation of the Green's function, which entails the calculation of the $N$-body ground state as well as eigenstates and associated energies of the $(N\pm1)$-body neighbors. 
While the former is accessed via variational techniques, the latter are constructed by means of the quantum subspace expansion method.
Different ansatzes for the ground-state wave function, originating from either classical or quantum approaches, are tested and compared to exact calculations. 
The resulting one-body Green's functions prove to be accurate approximations of the exact one for a large range of parameters, including across the normal-to-superfluid transition.
As a byproduct, this approach yields a good description of odd systems provided that the starting even system is well reproduced by the variational ansatz.
\end{abstract}

%\keywords{quantum computing, quantum algorithms}

\maketitle

\section{Introduction}

Many-body Green's functions constitute one of the methods of choice for the description of quantum $N$-body systems.
Current applications span several areas, ranging from condensed matter and quantum chemistry~\cite{Oni02} to nuclear physics, where the self-consistent Green's functions approach is used for both nuclear structure studies~\cite{Som20} and the investigation of the nuclear equation of state~\cite{Rio20}.
Exploratory calculations of nuclear reactions are also underway~\cite{Idi19,Vor24}.

One of the main advantages of the method is the ample flexibility in devising approximate yet accurate solutions to the many-body Schr\"{o}dinger equation.
Approximation schemes are usually handled at the level of the one-body self-energy, and take the form of either perturbative or non-perturbative truncated expansions that scale polynomially with the system size~\cite{Sch18,Bar17}. 
In nuclear physics, low-order expansion schemes are routinely employed and have proven successful in numerous nuclear structure applications~\cite{Som20,Som20b,Som21}.

In spite of the gentle scaling, current state-of-the-art nuclear structure calculations need to be performed on large supercomputers and are often based on heavily parallelized codes.
As a result, the associated computational costs limit extensions towards, e.g., heavy (and deformed) nuclei, high accuracy, uncertainty quantification, and application to nuclear reactions.
In this context, the emerging field of quantum simulations opens up new possibilities and, in the longer term, the promise of overcoming at least some of these computational limitations.

The application of quantum computers to many-body problems is gaining momentum~\cite{McA20,motta2022,Ayr23,fauseweh2024,wu2024}.
In nuclear physics, recent developments include the use of symmetry breaking and restoration techniques~\cite{Rui22,Lac23,Rui23,Rui24}, the implementation of shell-model calculations~
\cite{sarma2023,perez-obiol2023,nigro2024,yoshida2024,Bho24,yang2024,Cos24,Cos25,Yos25} and pionless effective field theory~\cite{Wat23,Gu25}, as well as applications of variational methods to light nuclei~\cite{dumitrescu2018,kiss2022,Car25}.

In related fields, mainly quantum chemistry and condensed matter, several works have addressed the calculation of Green's functions.
Recent proposals include imaginary- and real-time computations~\cite{End20,Sak22,Lib22,Dha24,Pic25}, methods based on Lanczos recursion~\cite{Bak21,Gre23} or local variational quantum compilation~\cite{Kan23}, and the use of equation-of-motion techniques~\cite{Riz22}.
To our knowledge, no quantum simulations of Green's functions specifically tailored to nuclear systems have been reported so far.

The present work aims at defining a specific strategy to compute Green's functions in (nuclear) many-body systems using a hybrid quantum-classical method.
The scheme, inspired by Ref.~\cite{Dha24}, consists of three steps.
First, the even-$N$ ground state is computed by means of a variational technique.
Second, states and energies in the odd neighbors are approximated via a quantum subspace expansion (QSE).
Finally, spectroscopic amplitudes and energy differences entering the Lehmann representation are computed to construct the one-body Green's function.
The approach is benchmarked on a system that is relevant for nuclear physics, namely the Richardson model~\cite{Ric64,Ric66,Duk04} whose pairing Hamiltonian generates strong superfluid correlations.
This model Hamiltonian is regularly used to test many-body methods on classical computers, including most recent applications employing neural networks \cite{Rig23}, matrix-product states \cite{Rau24} or Monte-Carlo estimates of Green's functions \cite{Bro25}.
Here, we compare different approaches on quantum computers that either exploit the symmetry-breaking / symmetry-conserving paradigm or seek a symmetry-conserving approximation to the correlated $N$-body ground state. 

The manuscript is organized as follows.
In Sec.~\ref{sec:formalism} we lay out the basic formalism and outline the hybrid quantum-classical strategy for the calculation of the Green's function.
Section~\ref{sec:application} is devoted to the application to the pairing Hamiltonian.
After a focus on the variational ground-state ansatzes employed in this study in Sec.~\ref{sec:variationalGS}, the calculation of the Green's function is addressed in Sec.~\ref{sec:GFcalculation}. 
A brief discussion on the description of odd systems is also carried out in Sec.~\ref{sec:odd}. 
Finally, conclusions and perspectives are addressed in Sec.~\ref{sec:conclusions}. 
Two appendices with useful demonstrations conclude the article.

\section{Calculation of many-body Green's functions}
\label{sec:formalism}

\subsection{Basic ingredients}
\label{sec:GF}

A set of $N$ particles governed by the two-body Hamiltonian
\begin{eqnarray}
H &=& \sum_{ij} \langle i | h | j \rangle a^\dagger_i a_j + \frac{1}{2} \sum_{ijkl} \langle ij | V| kl \rangle a^\dagger_i a^\dagger_j a_l  a_k  \, , \label{eq:hamilgen} 
\end{eqnarray}
is considered,
where $a^\dagger_i$ and $a_j$ represent a complete set of creation and annihilation operators associated to the single-particle basis $\{ | i \rangle \}_i$. 
Eigenstates and eigenenergies of the system are obtained upon solution of the $N$-body Schr\"{o}dinger equation
\begin{eqnarray}
H | \Psi_k^N \rangle = E_k^N | \Psi_k^N \rangle \, , 
\label{eq:schrodinger}     
\end{eqnarray}
with $| \Psi_0^N \rangle$ denoting the (normalized) $N$-body ground state.
The corresponding one-body Green's function is defined in real time as \cite{Fet71}
\begin{eqnarray}
G_{ij}(t,t') &\equiv& \langle \Psi_0^N |{\mathcal T} [a_i (t) a^\dagger_j(t')] | \Psi_0^N \rangle \, , 
\label{eq:gfrealtime}     
\end{eqnarray}
where ${\mathcal T}[.]$ is the time-ordering operator and the convention $\hbar =1$ is used.
Creation and annihilation operators appear in Heisenberg representation, e.g., $a^{\dagger}_i(t) \equiv U^\dagger(t) \, a^\dagger_i \, U(t)$, with $U(t) \equiv e^{-iHt}$. 
The one-body Green's function contains the most relevant physical information on the interacting many-body system.
In particular, its full knowledge allows the evaluation of ground-state expectation values of one-body operators and of the total energy via the Galitskii-Migdal-Koltun sum rule~\cite{Gal58,Kol72}.
In addition, when rewritten in the Lehmann representation (discussed below), it gives access to spectra of neighboring $N \pm 1$ systems.

The direct calculation of the Green's function via Eq.~\eqref{eq:gfrealtime} is in principle possible on a quantum computer.
This could be achieved in two steps, (i) a fermion--to--qubit mapping of the creation and annihilation operators entering the expectation value, and (ii) the translation of the evolution operators $U(t)$ into a quantum circuit using the Trotter-Suzuki method \cite{Tro59,McA20}.
Such a procedure, however, requires significant improvements to the quantum hardware (especially in terms of noise reduction).

A more viable path for implementation on current quantum devices exploits the spectral representation of the Green's function.
Provided that the system is time-translation invariant, i.e., $\forall t,t', G_{ij}(t,t') = G_{ij}(t-t',0)$, Eq.~\eqref{eq:gfrealtime} can be Fourier transformed to access the energy representation $G_{ij}(\omega)$.
Making use of eigenstates $\{| \Psi^{N\pm1}_k \rangle  \}_{k}$ and eigenenergies $\{E_{k}^{N\pm1} \}_{k}$ of the neighboring $N \pm 1$ systems, $G_{ij}(\omega)$ can be expressed in the so-called Lehmann representation~\cite{Leh54} as 
\begin{eqnarray}
G_{ij}(\omega) &=& \sum\limits_k \frac{\langle\Psi_0^N|a_i|\Psi_k^{N+1}\rangle \langle\Psi_k^{N+1}|a^\dagger_j|\Psi_0^N\rangle}{\omega - (E_k^{N+1} - E^N_0) + i\eta} \nonumber \\
&+& \sum\limits_{k'} \frac{\langle\Psi_0^N|a^\dagger_j|\Psi_{k'}^{N-1}\rangle \langle\Psi_{k'}^{N-1}|
a_i|\Psi_0^N\rangle}{\omega - (E_0^N - E_{k'}^{N-1}) - i\eta} \, ,
\label{eq:gfexact}
\end{eqnarray}
where $\eta > 0$ is a small parameter that shifts the poles away from the real axis.
This representation offers more flexibility than the expectation value~\eqref{eq:gfrealtime}.
In classical computations of nuclear or electronic systems, the knowledge of the exact energy dependence is usually exploited to derive an energy-independent eigenvalue problem that offers a convenient numerical implementation~\cite{Sch89,Som11}. 
Moreover, one can take advantage of the explicit presence of the exact eigenstates and eigenenergies in different particle-number sectors to devise suitable approximations for $G_{ij}$.
This strategy is followed in the present work for the calculation of the one-body Green's function on quantum platforms, as discussed in the following.

\subsection{General methodology for quantum simulations}
\label{sec:strategy}

%\begin{figure*}[htbp]
%%\includegraphics[width=0.8\linewidth]{figstrategy.png}
%\includegraphics[width=1.0\linewidth]{ComputMethodScheme.png}
%\centering
%\caption{Complete schematic view of the strategy we anticipate for computing Green's function using hybrid quantum-
%classical 
%computing. \samuel{If one keep this fig, I need to replace A by N to make the notations coherent.}{}}
%\label{fig:strategy}
%\end{figure*}

\begin{figure*}
\includegraphics[width=\linewidth]{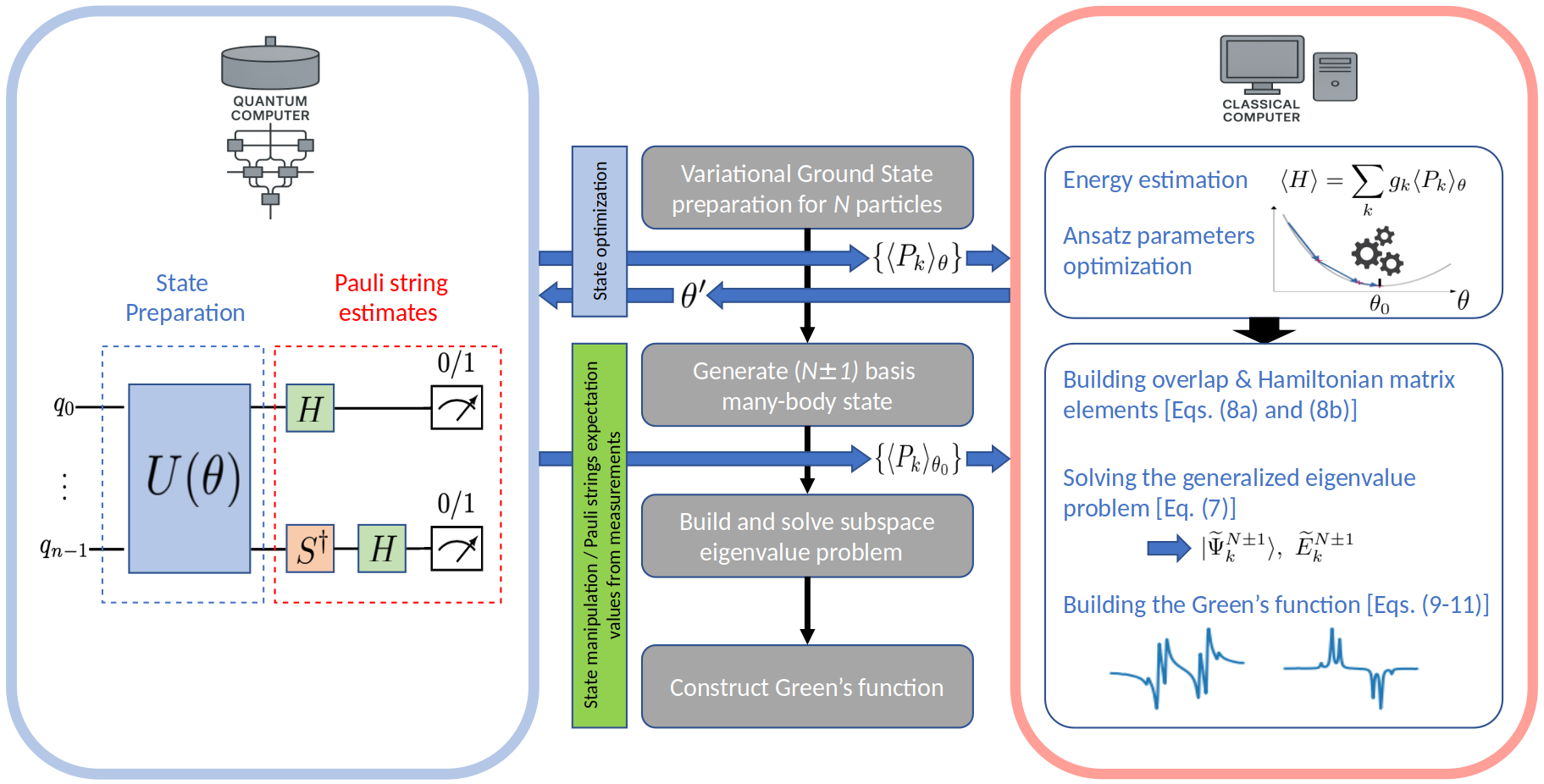}
\centering
\caption{Schematic representation of the procedure described in section \ref{sec:strategy}. The left part corresponds to the computations performed by the quantum processing unit (QPU), while the right part corresponds to those performed by the classical processing unit (CPU). Gray boxes indicate the main steps of the method's workflow.
The task of building the ansatz is performed on the QPU for a certain set of parameters denoted generically by $\theta$. Here, we denote generically by $U(\theta)$, the unitary operation preparing the state, such that the state before measurements reads as $| \Psi (\theta) \rangle = U(\theta) |0 \rangle^{\otimes n} $. In the present work, several choices of $U(\theta)$ are discussed. 
For given values of the parameters, Pauli strings expectation values, denoted generically as $\{\langle P_k \rangle \}$ are computed from measurements on the quantum register (see appendix~\ref{App:PauliTermsEval}).
The energy $\langle\Psi(\theta)|H|\Psi(\theta)\rangle$ is then reconstructed on the classical computer using the fact that the Hamiltonian writes as a linear combination of Pauli strings, $H=\sum_k g_k P_k$. The procedure is eventually iterated with new parameters $\theta'$ until convergence to an energy minimum, reached for a set of parameters denoted by $\theta_0$. The optimization task is performed on the CPU.
The lower part depicts how this ground state approximation is employed to evaluate the Hamiltonian and overlap matrix elements ${\cal H}^{N\pm 1}_{\beta \alpha}$ and ${\cal O}^{N\pm 1}_{\beta \alpha}$ defined in Eq.~\eqref{eq:hamil-ab}, which are then used to build two generalized eigenvalue problems whose solutions (obtained with a classical computer) constitute the building blocks for the Green's function. In this postprocessing step, the task devoted to the QPU is the preparation of the approximate ground state corresponding to $U(\theta_0)$ and the estimates of all necessary Pauli strings expectation values needed to compute ${\cal H}^{N\pm 1}_{\beta \alpha}$ and ${\cal O}^{N\pm 1}_{\beta \alpha}$ (see appendix~\ref{app:ExpectValuesEval}). The generalized eigenvalue problems and the reconstruction of the Green's function are finally performed on the classical computer.
}
\label{fig:Methodology}
\end{figure*}

A procedure to obtain the Green's function based on the Lehmann representation has been recently highlighted in Ref.~\cite{Dha24}.
The scheme makes use of hybrid quantum-classical techniques and builds upon the following steps:
\begin{enumerate}
    \item Given a parametrized circuit as the wave-function ansatz, the variational quantum eigensolver (VQE) hybrid method is used to obtain an approximation of the $N$-particle ground-state wave-function. This approximate ground state is denoted by $|\widetilde{\Psi}_0^N\rangle$ below. Note that only ansatzes possessing the correct particle number are considered here.
    \item To construct states with $N\pm 1$ particles, two pools of operators $\{ A^+_\alpha \}_{\alpha=1, \cdots, \Omega}$ and $\{ A^-_\alpha \}_{\alpha=1, \cdots, \Omega}$ are chosen. Each operator in the former (latter) has the effect of increasing (reducing) particle number by one unit when applied to the reference state $|\widetilde{\Psi}_0^N\rangle$, which results in two new sets of (eventually non-orthogonal) states
    \begin{subequations}
    \begin{eqnarray}
    \displaystyle
    \left\{|\phi_1^{N+1} \rangle, \cdots , |\phi_{\Omega} ^{N+1} \rangle \right\} 
   % \hspace{-0.1cm} &\equiv& \hspace{-0.1cm}
    &\equiv& 
       \left\{ A^+_1 , \cdots ,  
        A^+_{\Omega}  \right\} |\widetilde{\Psi}_0^N\rangle\, , \label{eq:phinp1} \\
    \displaystyle
       \left\{|\phi_1^{N-1} \rangle, \cdots , |\phi_{\Omega} ^{N-1} \rangle \right\}    
   % \hspace{-0.1cm} &\equiv& \hspace{-0.1cm}
    &\equiv& 
        \left\{ A^-_1, \cdots , A^-_{\Omega} \right\} |\widetilde{\Psi}_0^N\rangle\, . \label{eq:phinm1}
    \end{eqnarray}
    \end{subequations}
    Such states span reduced subspaces of the $(N+1)$- and $(N-1)$-body Hilbert spaces, respectively.
    \item Approximate eigenstates of the Hamiltonian for $N \pm 1$ particles 
    \begin{eqnarray}
     |\widetilde{\Psi}_k^{N\pm 1}\rangle &=& \sum_\alpha c^{N\pm 1}_\alpha(k) | \phi^{N\pm 1}_\alpha \rangle \, ,
     \label{eq:psiNpm1}
    \end{eqnarray}
    together with their corresponding eigenvalues $\widetilde{E}^{N\pm 1}_k$, can be obtained by solving a generalized eigenvalue problem in the two reduced subspaces defined above, which is written as
    \begin{equation}
        \sum_\beta c^{N\pm 1}_\alpha(k) {\cal H}^{N\pm 1}_{\beta \alpha} =  \widetilde{E}^{N\pm 1}_k \sum_\beta c^{N\pm 1}_\alpha(k) {\cal O}^{N\pm 1}_{\beta \alpha}  \, , 
        \label{eq:geneigen}
    \end{equation}
    with 
    \begin{subequations}    
    \label{eq:overlap-H}
    \begin{eqnarray}
        {\cal O}^{N\pm 1}_{\beta \alpha} %\hspace{-0.15cm} &\equiv& \hspace{-0.15cm} 
        &\equiv&
        \langle \phi^{N\pm 1}_{\beta} | \phi^{N\pm 1}_{\alpha}\rangle = \langle \widetilde{\Psi}_0^N | A^\mp_{\beta} A^\pm_{\alpha} | \widetilde{\Psi}_0^N \rangle \, , \label{eq:overlap} \\ 
        {\cal H}^{N\pm 1}_{\beta \alpha} %\hspace{-0.15cm} &\equiv& \hspace{-0.15cm} 
        &\equiv&
        \langle \phi^{N\pm 1}_{\beta} | H | \phi^{N\pm 1}_{\alpha}\rangle = \langle \widetilde{\Psi}_0^N | A^\mp_{\beta} H A^\pm_{\alpha} | \widetilde{\Psi}_0^N \rangle \, . \label{eq:hamil-ab}
    \end{eqnarray}
    \end{subequations}
    Following the hybrid quantum-classical paradigm, the numerical effort to build and solve Eq.~\eqref{eq:geneigen} is distributed between the quantum processor unit (QPU) and the classical one (CPU), as follows. 
    First, the $4 \Omega^2$ expectation values defined in Eqs.~\eqref{eq:overlap-H} are computed by the QPU: the operators $A^\mp_{\beta} A^\pm_{\alpha}$ and $A^\mp_{\beta} H A^\pm_{\alpha}$ are decomposed into combinations of Pauli terms, which are evaluated by measuring, after the suitable basis rotation, qubits prepared in a state encoding $|\widetilde{\Psi}_0^N \rangle$ (these decompositions and evaluations are detailed in appendix~\ref{app:ExpectValuesEval}).
    The obtained expectation values are then substituted into Eq.~(\ref{eq:geneigen}) that is solved on a CPU\footnote{The solution of the generalized eigenvalue problem is rather standard on classical computers and is made in two steps \cite{Rui21}. First, the overlap matrix is diagonalized to obtain an orthogonal basis for the reduced subspace of interest. Then, this orthogonal basis is employed to diagonalize the Hamiltonian matrix. Noteworthy, the diagonalization might lead to fewer states than in the set $\{| \phi^{N\pm 1}_\alpha \rangle\}$ if these states are not linearly independent. In practice, this situation does not lead to any specific difficulty.}. 
    Ultimately, the method provides approximations for $\Omega$ eigenstates and eigenvalues for $N\pm 1$ particles, $\{ | \widetilde{\Psi}_k^{N\pm1} \rangle , \widetilde{E}_k^{N \pm 1} \}_{k=0, \cdots, \Omega -1}$.
    While this resembles what is standardly referred to as configuration mixing in nuclear physics, in the quantum computing context it has been put forward as an efficient way to mitigate hardware noise and termed QSE~\cite{McC17,McC20b,Tak20,Yos22a,Jam22,Ume24,Gau24}. Notably, the QSE method has also been developed for nuclear physics model Hamiltonians such as the pairing model using the quantum Krylov technique \cite{Rui22} or in the Lipkin model using either the quantum computing extension of the generator coordinate method \cite{Bea24} or the so-called quantum equation of motion technique \cite{Hla22,Hla24}.
    
    \item Finally, an approximate Green's function is built using these eigenelements as
    \begin{eqnarray}
        G_{ij}(\omega) &=& \sum\limits_k \frac{\langle \widetilde \Psi_0^N|a_i|  \widetilde \Psi_k^{N+1}\rangle \langle  \widetilde \Psi_k^{N+1}|a^\dagger_j|\widetilde \Psi_0^N\rangle}{\omega - ( \widetilde E_k^{N+1} -  \widetilde E^N_0) + i\eta} \nonumber \\
        &+& \sum\limits_{k'} \frac{\langle \widetilde \Psi_0^N|a^\dagger_j| \widetilde \Psi_{k'}^{N-1}\rangle \langle  \widetilde \Psi_{k'}^{N-1}|
        a_i| \widetilde \Psi_0^N\rangle}{\omega - ( \widetilde E^N_0 -  \widetilde E_{k'}^{N-1}) - i\eta} \, .
        \label{eq:gfapproxlehmann}
    \end{eqnarray}
    The spectroscopic amplitudes appearing in the numerators in Eq.~\eqref{eq:gfapproxlehmann} can be obtained by combining expectation values $\langle \widetilde \Psi_0^N|a_i A^+_{\alpha} | \widetilde{\Psi}_0^N \rangle$ (which can be computed on the QPU) with the coefficients $c^{N\pm 1}_\alpha(k)$ (obtained by solving Eq.~(\ref{eq:geneigen})) as one can see by inserting Eq.~(\ref{eq:psiNpm1})
    \begin{eqnarray}
        \langle \widetilde \Psi_0^N|a_i| \widetilde\Psi_k^{N+1}\rangle &=& \sum\limits_{\alpha} c^{N+1}_\alpha(k) \langle \widetilde\Psi_0^N|a_i | \phi^{N+1}_\alpha \rangle \nonumber\\
        &=& \sum\limits_{\alpha} c^{N+1}_\alpha(k) \langle \widetilde \Psi_0^N|a_i A^+_{\alpha} | \widetilde{\Psi}_0^N \rangle \, ,
        \label{eq:compute_amplitudes}
    \end{eqnarray}
    and similarly with $\langle \widetilde \Psi_0^N|a^\dagger_j| \widetilde \Psi_{k'}^{N-1}\rangle$.
\end{enumerate}
This procedure is schematically summarized in Fig.\ref{fig:Methodology}.

Until now, we have not specified the operators $\{A^{\pm}_\alpha\}_\alpha$. 
A natural choice would be, for $A^{+}_\alpha$, to combine $(m+1)$ creation and $m$ annihilation operators to build $(m+1)$-particle--$m$-hole excitations as $\{ a^\dagger_i \}_{i}$, $\{ a^\dagger_i a^\dagger_j a_k \}_{i,j,k}$, \dots, with $m=0,\dots,N$ (and similarly for $A^{-}_\alpha$).
In the present study, we simply consider $\{ A^+_{\alpha} \}_\alpha = \{ a^{\dag}_i \}_{i}$ and $\{ A^-_{\alpha} \}_\alpha = \{ a_i \}_{i}$, so that the expectation values entering Eq.~\eqref{eq:compute_amplitudes} read as
\begin{subequations}    
\begin{align}
    \langle \widetilde \Psi_0^N|a_{j} A^+_{i} | \widetilde{\Psi}_0^N \rangle &= \langle \widetilde \Psi_0^N|A^-_{j} a^{\dag}_{i} | \widetilde{\Psi}_0^N \rangle = {\cal O}^{N+1}_{ji} \, , \\
    \langle \widetilde \Psi_0^N|a^{\dag}_{j} A^-_{i} | \widetilde{\Psi}_0^N \rangle &= \langle \widetilde \Psi_0^N|A^+_{j} a_{i} | \widetilde{\Psi}_0^N \rangle = {\cal O}^{N-1}_{ji} \, .
\end{align}
\end{subequations}    
This choice leads to the benefit that overlaps from step 3 can be reutilized, and the QPU evaluation of the amplitudes in step 4 can be skipped, which results in a significant computational gain.

In summary, in the strategy depicted above, the manipulation of the complicated many-body states in Fock space is assigned to the QPU. 
This manipulation includes the circuit used to build the ansatz for the many-body wave function, and the calculation of expectation values of operators. 
All other tasks are performed on a classical machine, including the ansatz optimization, the subspace diagonalization, and the calculation of the one-body Green's function from the expectation values.
Below, we apply this approach to small superfluid systems %\samuel{to a small superfluid system ? (singular ?)}{} 
and evaluate its performance using noiseless emulators of quantum computers.

\section{Application to the pairing Hamiltonian}
\label{sec:application}

\subsection{The pairing model}

We consider here a system of $N$ particles interacting through a pairing-type Hamiltonian with constant 
two-body coupling, which reads as
\begin{eqnarray}
    H(g) = \sum\limits_{k=1}^{D} \varepsilon_k (a_k^{\dag}a_k + a_{\bar{k}}^{\dag}a_{\bar{k}}) 
    - g\sum\limits_{k\neq l} a_k^{\dag} a_{\bar{k}}^{\dag} a_{\bar{l}}a_l \, ,
    \label{eq:Hpairing}
\end{eqnarray}
where $\varepsilon_k$ represent single-particle energies, and $g$ is the interaction strength.
The single-particle state labeled by $\bar k$ denotes the time-reversed state of $k$, with the two assumed to be degenerate in energy.
Equidistant doubly-degenerate energy levels are considered, i.e., $\varepsilon_k = k \Delta E$ with $k=1, \dots, D$. 
Note that the corresponding number of single-particle levels is $n_{\rm sp} = 2D$. 
In the following, energies will be given in units of $\Delta E$.  

The Hamiltonian (\ref{eq:Hpairing}) is widely used to describe small superfluid systems \cite{Von01,Duk04} or to have a schematic account of pairing effects in atomic nuclei \cite{Zel03,Bri05}. 
Despite its simplicity, it represents an archetype of a non-perturbative many-body problem where it can be useful to break the $U(1)$ symmetry associated with particle number, a technique that is extensively used nowadays in the nuclear many-body problem \cite{Rin80,Bla86,Ben03}.
While a generalization of many-body Green's functions to a particle-number breaking framework exists~\cite{Som11} and is routinely employed in nuclear structure calculations~\cite{Som20}, such an approach is not considered here.

The pairing model has been recently used as a benchmark for quantum algorithms and quantum ansatzes \cite{Lac20,Kha21,Rui21,Rui22,Rui23,Lac23,Rui24} (see also the pioneering work of Ref.~\cite{Ovr03,Ovr07}). 
Two strategies have been employed so far to obtain accurate quantum ansatzes. 
The first one follows the two-step procedure consisting of (i) preparing a symmetry-breaking state that grasps strong collective correlations, followed by (ii) a symmetry-restoration approach. 
More recently, variational symmetry-preserving ansatzes have also been successfully applied to the pairing problems~\cite{Kha23,Zha24}. 
Both schemes are tested in the present work.

Below, after a brief description of the full configuration interaction method on classical computers that will serve us as a reference calculation, we illustrate how the strategy discussed in section \ref{sec:strategy} is implemented for the pairing Hamiltonian model.

\subsection{Exact Green's functions on classical computers}

For small system sizes, all quantities entering Eq.~\eqref{eq:gfexact}, i.e., the $N$-body ground state, the ground-state energy $E_0^N$, as well as $(N\pm1)$-body eigenstates and associated eigenenergies, can be computed exactly via full configuration interaction (FCI) on a classical machine.
\begin{figure*}
\includegraphics[width=\linewidth]{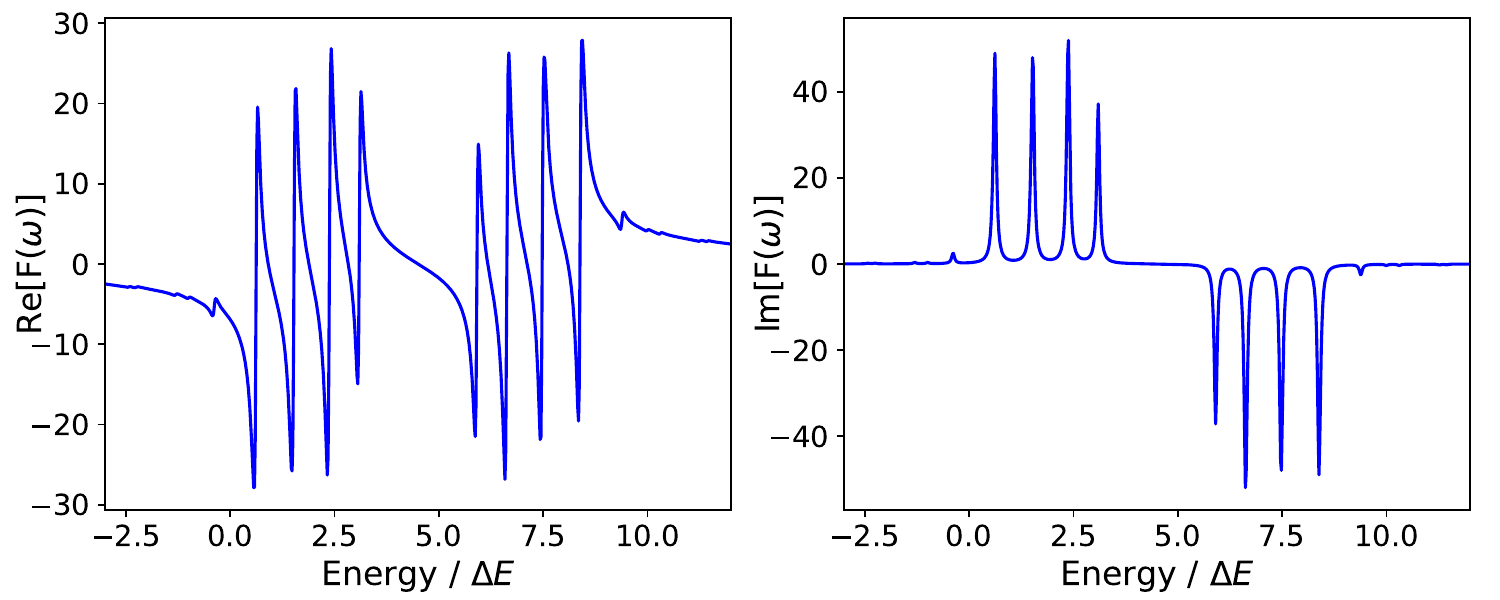}
\centering
\caption{Real (left) and imaginary (right) part of the function  (\ref{eq:traceG}) for a pairing system of $N = 8$ particles on $D = 8$ doubly-degenerated levels with coupling strength $g=0.6$. The curves represent the exact result obtained via FCI on a classical computer, with a width (see Eq.~\eqref{eq:gfexact}) set to $\eta = 0.04$.}
\label{fig:Exact_GFT}
\end{figure*}
For this purpose, we consider a basis of the $2^{2D}$-dimensional Fock space denoted by 
\begin{eqnarray}
\{| \phi_L \rangle\}_{L=0, \cdots,  2^{2D}-1} \, .
\end{eqnarray}
In order to get closer to a quantum computing notation, one can write the basis as
\begin{eqnarray}
\{| [L] \rangle\}_{L=0, \cdots,  2^{2D}-1} \, , 
\end{eqnarray}
where
\begin{eqnarray}
    [L] = b^L_{2D-1} \cdots b^L_0
    \label{eq:binnotation}
\end{eqnarray}
denotes the binary number associated with $L$. As in the quantum computing case (see below), one can relate the coefficient $b_i$ to the occupation ($b_i =1$) or not ($b_i =0$) of the single-particle state $i$.

On classical computers, one can significantly reduce the numerical effort by exploiting the symmetries of the problem. 
First, if one is interested in states with a fixed number of particles $N$, one can diagonalize the Hamiltonian in the subspace of wave functions that have $N$ occupied states. 
Such wave functions correspond to states where the Hamming weight of the binary string $[L]$, i.e., the number of $1$, is exactly $N$. 
The size of this subspace is $\binom{2D}{N}$. 
Moreover we focus, in the following, on systems with an even number of particles. 
In this case, the ground state is characterized by seniority $\nu = 0$, i.e., there are no unpaired particles or, equivalently, if the single-particle state $k$ is occupied, $\bar k$ is also occupied.
Using this property, one further reduces the size of the relevant subspace to $\binom{D}{N/2}$.
For instance, if we consider $8$ particles on $8$ (doubly degenerated) levels, the full Fock space has a size of $2^{16} = 65536$ states while it reduces to $\binom{8}{4} = 70$ once using only states with good particle number and $\nu=0$. For further details on the use of the seniority concept for Hilbert-space reduction, we refer the reader to the extensive discussions in \cite{Rac43,Des63,Zel03,Bri05}.

To obtain the eigenstates with $N\pm 1$ particles, one can reduce the dimension in a similar way. 
Since the $N$-body ground state has seniority zero, from expression (\ref{eq:gfexact}) it follows that only $(N\pm 1)$-body states with  $\nu =1$ will contribute to the Green's function.
This reduces the size of the relevant subspaces to $(2D - N + 2) \binom{D}{(N-2)/2}$ and $(2D - N) \binom{D}{N/2}$ for $N-1$ and $N+1$ particles respectively.
In addition, the Hamiltonian (\ref{eq:Hpairing}) has a block-diagonal structure in these spaces, since it does not couple states in which the unpaired particle is in different energy levels.
This leads to a further simplification of the numerical algorithm.

In practice, for a given $N$, we performed three FCI calculations related to $N-1$, $N$, and $N+1$ particles, with states written in the Fock space as described above. 
The exact one-body Green's function can be computed using Eq.~(\ref{eq:gfexact}). 
As an example, we show in Fig.~\ref{fig:Exact_GFT} the results for $N = 8$ particles on $D = 8$ doubly-degenerated levels (half-filling). 
Specifically, the real and imaginary part of the function
\begin{eqnarray}
    F(\omega) &=& {\rm Tr}[G(\omega)]
\label{eq:traceG}
\end{eqnarray}
are displayed in different panels that will serve as reference calculations for testing quantum algorithms.

\subsection{Encoding the pairing problem on digital quantum computers}
\label{subsec_JW}

Let us now address the solution of the pairing model following the hybrid quantum-classical procedure outlined in Sec.~\ref{sec:strategy}.
First, a method to encode fermionic problems on a qubit register has to be chosen. 
Here, we use the traditional Jordan-Wigner  transformation (JWT) \cite{Jor28, Lie61}, where each qubit corresponds to one single-particle state. 
Qubits are labeled by $\alpha=0, \cdots, n_{\rm sp}-1$. 
Standard notations $(X_\alpha, Y_\alpha, Z_\alpha)$ are used for the three Pauli matrices acting on the qubit $\alpha$. 
%
%In the following, for a given operator $O$ acting in Fock space, we use the notation ${\rm JW}[O]$ for the equivalent JW-transformed operator acting on qubits.
To perform the JWT, one needs to specify how single particle modes are ordered on qubits. 
As underlined in Ref.~\cite{Lac20}, in the present context it is convenient to order them in consecutive time-reversed pairs with $(i,\bar i) = (\alpha, \alpha+1)$ where $\alpha=2i$ are qubit labels and $i=0, \cdots, D-1$.   

The building blocks of the JWT technique are the transformed creation/annihilation operators
\begin{subequations}    
    \label{EqJWT}
\begin{eqnarray}
    a_\alpha & \longrightarrow &  \left[ \prod\limits_{0\leq p < \alpha }\!\!(-Z_p)\right] \sigma^{+}_\alpha,  \\ 
    a^\dagger_\alpha & \longrightarrow &  \left[\prod\limits_{0\leq p < \alpha }\!\!(-Z_p)\right] \sigma^{-}_\alpha \, ,
\end{eqnarray}
\end{subequations}    
where $\sigma^{\pm}_\alpha \equiv \frac{1}{2}(X_\alpha \pm iY_\alpha)$.
The qubit representation of any $K$-body operator, with $K=1,\dots,N$, can be then obtained starting from its Fock-space expression and using the fact that Eqs.~\eqref{EqJWT} are linear and preserve operator products, tensor products, and hermitian conjugation~\cite{McC20,Fan19,Cao19,McA20,Bau20,Ayr23}.
Illustration of observables expressions obtained using JWT are given in appendices~\ref{app:CoherenceJWTpairs} (for Hamiltonian~(\ref{eq:Hpairing})) and~\ref{app:PauliOpExpansion_HandO} (for expectation values~(\ref{eq:overlap-H})).
For further details on the pairing problem using particle-to-qubit encoding see Refs.~\cite{Ovr03,Ovr07,Lac20}.

As argued previously, one can search for the $N$-body ground state in the subspace with no broken pairs. As was first noted in Ref.~\cite{Kha21}, assuming that the interaction is dominated by particles forming quasi-bosons, it is convenient to reduce the number of qubits required to describe them. Indeed, for even systems, one can directly encode the pair occupation on qubits, reducing the qubit number by a factor of two compared to the case where particle occupations are encoded.
As further discussed in appendix~\ref{app:CoherenceJWTpairs}, this leads to the simplification that a pair of particles can be encoded into a single qubit. This approach is called hereafter pair-to-qubit encoding. To do so, one can rewrite the pairing Hamiltonian~\eqref{eq:Hpairing} (in the $\nu=0$ subspace) as
\begin{equation} \label{eq:Hpairing_pairs}
    H(g) = \sum\limits_{k=1}^{D} 2\varepsilon_k P_k^{\dag}P_k - g\sum\limits_{k\neq l} P_k^{\dag} P_l \, , 
\end{equation}
where $P_k^{\dag} \equiv a_k^{\dag} a_{\bar{k}}^{\dag}$ and $P_k \equiv a_{\bar{k}} a_k$ are the pair creation and annihilation operators. 
%\textcolor{blue}{
The above pairing Hamiltonian has already been 
used to benchmark quantum algorithms \cite{Ovr03,Ovr07,Lac20,Kha21,Rui21,Rui22}. Recently, starting from a shell model Hamiltonian, a generalized pairing-like Hamiltonian that does not assume a constant two-body interaction was also introduced in \cite{yoshida2024,Yos25,Cos25}. 
%\samuel{The above blue sentences seems redundant to me, especially with the beginning of the paragraph (above Eq.18). I would replace it with a sentence like "The coherence of this pair-to-qubit encoding with the particle-to-qubit one is discussed in appendix A. For more details, we refer to ref~\cite{yoshida2024}." Do we really need to put 3 references instead of one ? As a reader I would prefer one reference, if it is sufficient, to save time.}{}
In the present work, we use the pair-to-qubit encoding scheme during the variational optimization step to prepare, for instance, the ground state of even-particle systems.

Nevertheless, the use of the Lehmann representation requires building a wave function having either $N$ or $N \pm 1$ particles. 
Therefore, we also need to consider odd systems where at least one pair is broken. 
For this reason, we are forced to use the standard encoding as well, called hereafter particle-to-qubit encoding and given in Eqs.~\eqref{EqJWT}. 
Both types of encoding, as well as their connection, are discussed in appendix~\ref{app:CoherenceJWTpairs}, including the resulting form of the Hamiltonian once 
encoded on the qubit register.

\color{black}
Notably, if we use the pair-to-qubit encoding to compute the ground state of the system with even $N$, the calculation of Green's functions requires a conversion of qubit states back to the usual JWT encoding. 
This can be done in a simple way, as follows.
Starting from a good approximation of the ground state prepared on a $D$-qubit quantum circuit in pair encoding, one can (i) add, for each qubit $q_k$, another qubit $q_{\bar{k}}$ initialized in $0$, then (ii) entangle them with a CNOT gate controlled by $q_k$. 
% In this way, (the measurement of) $q_k$ and $q_{\bar{k}}$ would correspond to $0$ or $2$ particles in the same level $k$.
The resulting $2D$-qubit quantum circuit is thus ready to be used to compute the matrix elements~\eqref{eq:overlap-H}, eventually providing the Green's function's building blocks.

\subsection{Variational ground-state ansatz}
\label{sec:variationalGS}

In this work, we compare several variational ansatzes for the $N$-body ground state.
Specifically, two strategies are employed to construct approximate ground-state wave functions $| \widetilde{\Psi}_0^N \rangle $. 
The first one is based on BCS theory, a method that is traditionally used in nuclear physics to obtain an expressive ground-state ansatz.
This approach exploits the breaking and restoration of the $U(1)$ symmetry associated to particle-number conservation~\cite{Rin80,Bla86}.
The second strategy makes use of standard quantum computing techniques that directly target and optimize a wave function within the proper $N$-particle subspace. 
Technical details for the different ansatzes are discussed below. 

\subsubsection{Particle-number projected BCS state }

Let us consider the BCS state
\begin{eqnarray}
    | \Phi \rangle &\equiv& \prod_{i=0}^{q-1}  \left(u_i + v_i a^\dagger_i a^\dagger_{\bar i} \right) | - \rangle \, ,
    \label{eq:bcswf}
\end{eqnarray}
where $| - \rangle$ is the particle vacuum.
The real parameters $(u_i,v_i)$ determine the Bogolyubov transformation linking the single-particle and BCS quasiparticle creation/annihilation operators \cite{Bri05}. 
It is easy to see that the state~\eqref{eq:bcswf} has components associated with different particle numbers.
One can obtain a state with the correct number of particles $N$ by using projection techniques, which gives the ansatz
\begin{eqnarray}
    | \widetilde{\Psi}_0^N \rangle &=& \frac{1}{\sqrt{\langle \Phi |P_N| \Phi \rangle }} P_N | \Phi \rangle \,  , \label{eq:prostate}
\end{eqnarray}
where $P_N$ is the projector on the subspace of wave functions with $N$ particles. 
Both BCS and projection techniques are rather standard and, for more details, we refer the reader to Refs.~\cite{Rin80,Bla86,Ben03,Bri05}. 

BCS-type states can be directly encoded on quantum computers and optimized through standard VQE techniques \cite{Ovr03,Ovr07,Lac20}. 
Several methods have also been proposed to restore symmetries, in particular particle number \cite{Lac20,Kha21,Rui21,Rui22,Lac23,Rui24,Rui23}.
Recently, an alternative approach based on the binary tree state technique has been suggested to directly obtain the projected BCS state without going through the symmetry-breaking step \cite{Kha23}. 
Here, we use a more direct method that can be applied if the number of qubits is moderate. 
We first prepare the BCS state on a classical computer by solving the associated gap equation with the constraint that the average particle number matches $N$. 
After this procedure, we obtain the set of $(u_i,v_i)$ minimizing the BCS energy. 
Then we build the projected state by simply selecting the components with the right particle number and normalizing the resulting state.
This scheme goes under the name of \textit{projection after variation} (PAV).
We also implemented the so-called \textit{variation after projection} (VAP) alternative, in which the projection is performed before variation of the $(u_i,v_i)$ coefficients. 
Both PAV- and VAP-BCS can be regarded as simple yet accurate procedures, easily implementable on classical computers, allowing for the account of superfluid effects in small systems.
These methods will serve as a reference in the following.

%\begin{figure}[h]
%\includegraphics[width=8.6cm]{Fidelities-Energies.pdf}
%\centering
%\caption{\samuel{Now the legend would be something like : (a) Fidelity computed between the exact ground state and the ones obtained with BCS (red), projected BCS (orange), and ADAPT-VQE with 4 iterations (green), as a function of the coupling strength $g/\Delta E$. (b) Energy of the corresponding states compared to the exact one (black). Results are shown here for $N=4$ particles on 4 doubly-degenerated levels.}{} (a) Comparison of the exact GS energy (solid line), BCS energy (dashed line) and Projected BCS 
%energy as a function of the coupling strength $g/\Delta E$. Panel (b): Corresponding Error on the corelation energy defined as the total energy minus the energy of the Hartree-Fock state. Results are shown here for $N=4$ particles on 4 doubly-degenerated levels.}
%\label{fig:E_exact_BCS_projBCS}
%\end{figure}

\subsubsection{Particle-number conserving ADAPT-VQE state}

We now present an alternative set of quantum computing ansatzes for the preparation of the pairing Hamiltonian ground state.
These ansatzes are based on or inspired by the adaptive derivative-assembled pseudo-Trotter ansatz variational quantum eigensolver (ADAPT-VQE) approach~\cite{Gri19}.
In Ref.~\cite{Zha24}, it was shown that such an approach works surprisingly well for the pairing problem without resorting to the breaking of particle-number symmetry.
The standard ADAPT-VQE, as well as several improved versions, are well documented in the literature, see e.g. Ref.~\cite{Gri19,Tan21,Yao21,Yor21,Zha21,Sma21,Hai22,Liu22,Yor22,Ber23,Fen23a,Gri23,Ana24}.
Here, we only review its minimal ingredients that are useful for the present discussion.

The method starts with a set of operators acting on many-body states, denoted by $\{ G_\alpha\}$.
Given these operators, an ansatz for the ground state is built up iteratively such that 
after $n$ steps, one has
\begin{eqnarray}
| \varphi_n \rangle &=& \prod_{i=1}^{n} e^{i \theta_i G_{\alpha_i}} | \varphi_0 \rangle \, ,
\label{eq:trialadapt}
\end{eqnarray} 
with $\theta_i$ representing the variational parameters.
In the simplest implementation, iterations proceed as follows
\begin{enumerate}
    \item $| \varphi_0 \rangle$ is a chosen state that serves as a seed to initiate the iterative process. Here we use the simple state where only the lowest energy levels are occupied, which in term of qubits reads as $| \varphi_0 \rangle = | 1 \cdots 1 0 \cdots 0 \rangle$.
    \item At each iteration $n \geq 1$, an operator is selected under the criterion that the energy gradient 
        \begin{eqnarray}
        \left. \frac{\partial E_n }{\partial \theta_n}\right|_{\theta_n=0} &=& i  \langle \varphi_{n-1} |\left[ H, G_{\alpha_n} \right]| \varphi_{n-1} \rangle \, ,
        \label{eq:gradient}
    \end{eqnarray}
    with
    \begin{eqnarray}
    E_n \equiv \langle \varphi_{n} | H | \varphi_{n} \rangle \, ,
    \end{eqnarray}
    is extremized. A new trial state is then built following Eq.~\eqref{eq:trialadapt}.
    Finally, in order to speed up the convergence, the full set of parameters $(\theta_1, \cdots , \theta_n)$ is re-optimized by minimizing the energy $E_n$.
    \item The iterative procedure is stopped when the gain in energy between two steps is below a given threshold.
\end{enumerate}
  
Several different choices can be made for the set of operators $\{ G_\alpha\}$.
Following Ref.~\cite{Zha24}, we first tested the so-called single qubit excitation-based pool (QEB-Pool) proposed in Ref.~\cite{Yor21}.
This set has the advantage of performing unitary operations that preserve particle number. 
Therefore, if the initial state $| \varphi_0 \rangle$ is an eigenstate of the particle-number operator, the various states generated during the ADAPT-VQE process will remain in the same symmetry block. 
The above iterative procedure, selecting operators based on a local maximization of energy gradients around the state $| \varphi_n \rangle$, will be called hereafter ``ADAPT-St", where 'St' stands for standard implementation. 

As pointed out in Ref.~\cite{Fen23b}, it might be convenient to use a more global criterion, taking advantage of the fact that the set of functions defined as
\begin{eqnarray}
E_n(\theta_n) &\equiv& \langle \varphi_{n-1} | e^{-i\theta_n G_n}H e^{i\theta_n G_n} | \varphi_{n-1} \rangle
\label{Etheta_in_AdaptVQE}
\end{eqnarray}
can often be written as a simple expression in terms of the angle $\theta_n$. 
Specifically, for the operator pool considered here, one can always write (see appendix~\ref{app:DemoEq23})
\begin{equation}
E_n(\theta_n) = C_0 + C_1\!\cos(\theta_n \!+\! \gamma_1) + C_2\!\cos(2\theta_n \!+\! \gamma_2) \, ,
\label{EnergySineForm}
\end{equation}
with only five parameters ($C_0,C_1,C_2,\gamma_1,\gamma_2$) that depend on the wavefunction and the operators themselves.
Therefore, in practice, the knowledge of $E_n(\theta_n)$ for five different values of $\theta_n$ is sufficient to infer the whole energy curve (see appendix~\ref{app:FixingParameters}).
This allows us to replace the local criterion based on the gradient, Eq.~\eqref{eq:gradient}, with a global one that selects the pool operator leading to the lowest possible point on all energy curves Eq.~\eqref{EnergySineForm}. 
The ADAPT-VQE scheme using this alternative criterion is labeled ``ADAPT-Min". 
Note that this algorithm is slightly more costly than ADAPT-St in selection of operators, since five evaluations per operator are required (vs a single evaluation per operator in ADAPT-St).
However, one has to keep in mind that the operator with the largest gradient at $\theta = 0$ is not always the one leading to the smaller energy minimum.

\begin{figure}
% (a)
% \begin{quantikz}[column sep = 7]
% \lstick{$i$}&&\targ{}\gategroup[2,steps=7,style={dashed,rounded
% corners,inner sep=3pt},label style={label position=below,anchor=north,yshift=-0.2cm}]{$B_{i,i+1}(\theta)$}&&&\ctrl{1}&&&\targ{}&& \\
% \lstick{$i\!+\!1$}&&\ctrl{-1}&\gate{Z}&\gate{R_Y(-\theta)}&\targ{}&\gate{R_Y(\theta)}&\gate{Z}&\ctrl{-1}&&
% \end{quantikz}
% %\includegraphics[width=8.0cm]{GardBlock.png}

% \vspace{16pt}

% (b)
% \begin{quantikz}[column sep = 6, row sep={0.8cm,between origins}]
% \lstick{$0$}&\gate{X}&&&&\gate[2][0.8cm]{B(\theta_9)}&& \\
% \lstick{$1$}&\gate{X}&&&\gate[2][0.8cm]{B(\theta_6)}&&\gate[2][0.8cm]{B(\theta_{12})}& \\
% \lstick{$2$}&\gate{X}&&\gate[2][0.8cm]{B(\theta_3)}&&\gate[2][0.8cm]{B(\theta_8)}&& \\
% \lstick{$3$}&\gate{X}&\gate[2][0.8cm]{B(\theta_1)}&&\gate[2][0.8cm]{B(\theta_5)}&&\gate[2][0.8cm]{B(\theta_{11})}& \\
% \lstick{$4$}&&&\gate[2][0.8cm]{B(\theta_2)}&&\gate[2][0.8cm]{B(\theta_7)}&& \\
% \lstick{$5$}&&&&\gate[2][0.8cm]{B(\theta_4)}&&\gate[2][0.8cm]{B(\theta_{10})}& \\
% \lstick{$6$}&&&&&&& \\
% \end{quantikz}
% \centering
% \vspace{-15pt}
\includegraphics[width=1.0\linewidth]{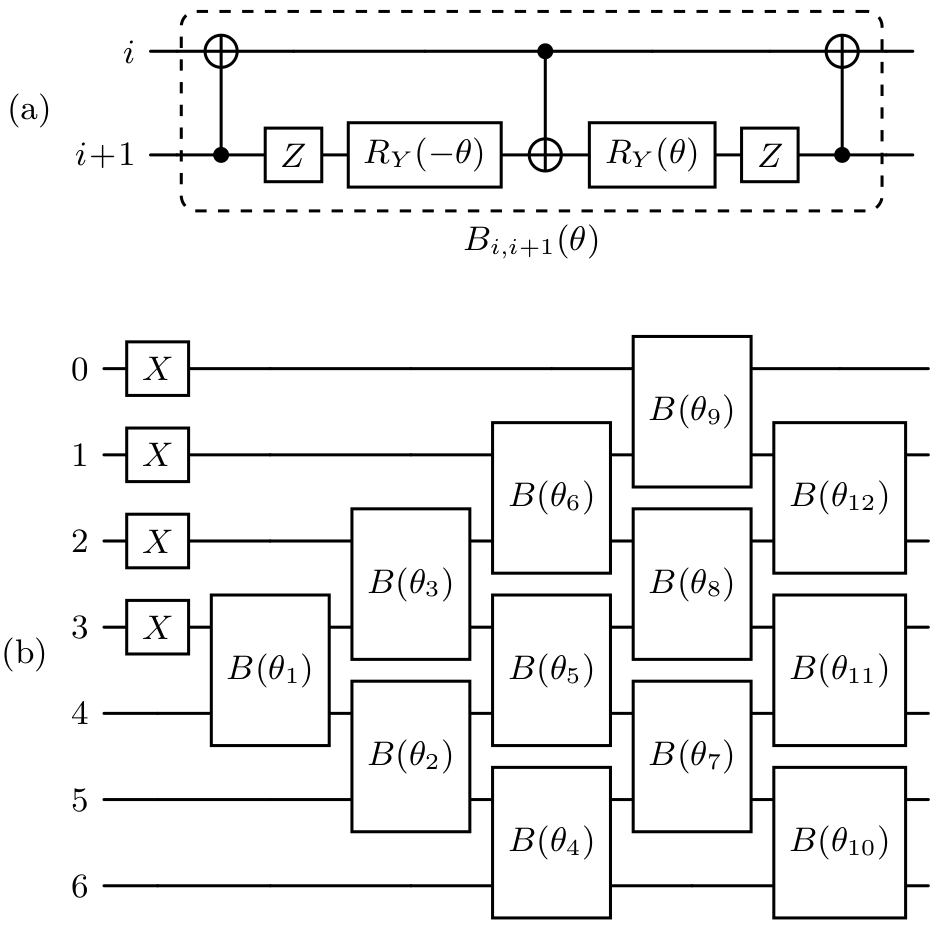}
\caption{Circuit corresponding to the operator $B_{i,i+1}(\theta)$ used in the ADAPT-Fix approach (a), and example of circuit using these blocks (b). In the latter, the angles indices indicate the order in which the operator are placed in this approach (other indices have been dropped here for clarity).} 
%\denis{A faire avec le package quantikz} \samuel{Ok I did it below.}{} }
\label{fig:GardBlock}
\end{figure}

Finally, we tested a class of ansatzes inspired from the iterative process proposed in Ref.~\cite{Gar20}. 
Here, the trial wave function is updated according to
\begin{eqnarray}
    | \varphi_n \rangle &=& B_{i_n , i_n +1}(\theta_n) | \varphi_{n-1} \rangle \, ,
    \label{eq:gardupdate}
\end{eqnarray}
where $B_{i,i+1}(\theta)$ is a two-qubit operator acting on qubits $(i,i+1)$. 
This operator, depicted in Fig.~\ref{fig:GardBlock}(a), is a simplified version of the more general operator proposed in Ref.~\cite{Gar20}. Note that, here we used the most general circuit able to describe any rotation of the $O(4)$ group. As 
illustrated in Fig.~\ref{fig:GardBlock}(a), the circuit requires three CNOT gates. Without lost of generalities, one can eventually restrict the rotations in the $SO(4)$ group, reducing the required number of CNOTs to two \cite{Vat04}.
As for the ADAPT-VQE variants described above, at each iteration one among the ($D-1$, in this case) possible operators is added, its corresponding angle is optimized, and then the whole set of angles $\{ \theta_1, \cdots, \theta_n \}$ is re-optimized. 
We have tested the operator selection within this operator pool following the same local and global criteria discussed before.
However, we have found that both criteria are unsatisfactory, often leading to plateaus and/or oscillatory behavior. 
One reason is that there exists no $\theta$ for which $B_{i,i+1}(\theta)$ is the identity operator, which invalidates the use of the gradient method.
Additionally, as soon as qubits $i$ and $i+1$ are entangled with other qubits, one can reach situations where, whatever $\theta$, the energy of $B_{i, i+1}(\theta) |\psi_{n-1}\rangle$ is strictly higher than the energy of $|\psi_{n-1}\rangle$. 
As a consequence, it is not guaranteed that the energy will decrease or at least remain equal at each step.

We have thus decided to select the operators in Eq.~\eqref{eq:gardupdate} according to a predefined order, similar (but not identical) to the one suggested in Ref.~\cite{Gar20}. 
Starting from the lowest energy Hartree-Fock state $| \varphi_0 \rangle = | 1 \cdots 1 0 \cdots 0 \rangle$, encoded in the quantum circuit by X-gates applied on qubits corresponding to occupied levels, the first $B_{i,i+1}(\theta)$ operator is applied on the two qubits $i$ and $i+1$ corresponding respectively to the highest occupied level and the lowest unoccupied one. Then the operators are placed one after the other in successive columns to add more and more correlations, as illustrated on Fig.~\ref{fig:GardBlock}(b).
In the following, this scheme is referred to as ``ADAPT-Fix''.

\begin{figure*}
\includegraphics[width=0.95\linewidth]{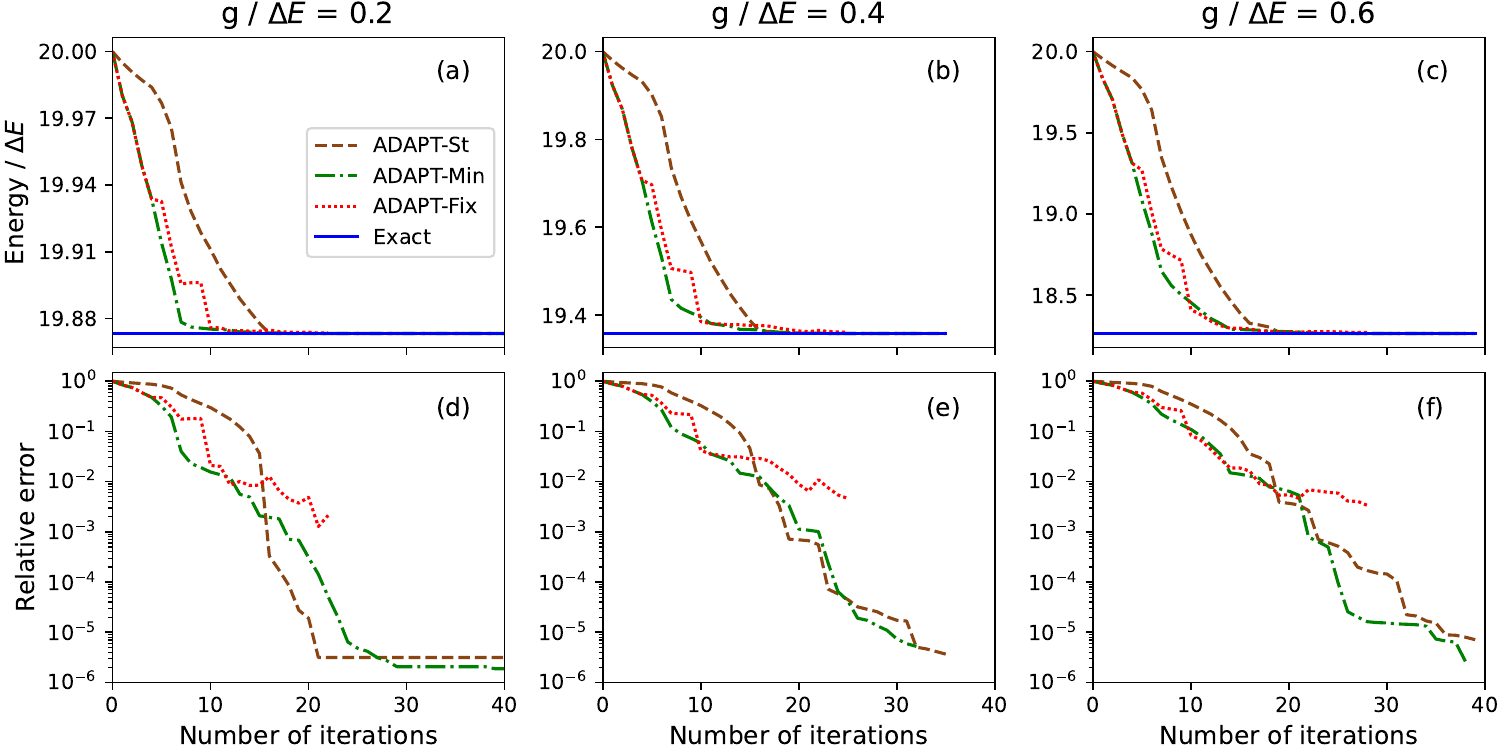}
\centering
\caption{Ground-state energy as a function of the number of iterations computed with the adaptative methods described in the text, for $N=8$ particles in $D=8$ levels and for $g/ \Delta E = 0.2$ (a), $0.4$ (b) and $0.6$ (c).
Note that, in the top panels, the range covered by the vertical axis varies to help visualize the convergence to the exact ground-state energy (represented by a blue horizontal line).
The corresponding relative errors defined in Eq.~\eqref{eq:eer} are shown in panels (d), (e), and (f), respectively.
The computations were performed on a classical machine simulating a quantum device, with a maximum allowed wall time of 24 hours. In the present figure, all iterations completed within this wall time are shown. As a consequence of this practical constraint, it might happen that the curves have not all reached full convergence.}
\label{fig:Energies_Nl8_Np4_g0.2to0.6}
\end{figure*}

\subsubsection{Results}

We have tested these adaptive methods in the context of the pairing model for different particle numbers $N$, energy levels $D$, and interaction strengths $g$. 
Calculations have been performed using the quantum emulation software QISKIT~\cite{qiskit}.
Figure~\ref{fig:Energies_Nl8_Np4_g0.2to0.6} shows the resulting ground-state energies for the case $N=D=8$ (i.e., half filling) and for three values of $g$ as a function of the iterations in the ADAPT-VQE algorithm.
In addition to total energies (top panels), in the bottom panels we display relative errors defined as
\begin{eqnarray}
   \delta E &=& \frac{E-E_{\rm exact}}{E_{\rm HF} - E_{\rm exact}}  \, ,
   \label{eq:eer}
\end{eqnarray} 
where $E_{\rm HF}$ and $E_{\rm exact}$ are respectively the Hartree-Fock and the exact ground state energies. 
While all algorithms yield a quick convergence towards the exact value, one notices that the global criterion (ADAPT-Min) outperforms the local one (ADAPT-St) in the first iterations.
However, as it can be seen in the relative errors, around 15 iterations, the ADAPT-St convergence eventually accelerates and becomes equivalent to ADAPT-Min in the long run.
The ADAPT-Fix scheme produces more irregular results.
Despite its fixed order, it converges as fast as ADAPT-Min for large couplings. 
In some cases, we even observed that it starts outperforming the two other procedures (e.g., around 15 iterations, for a system with $N=D=10$ and $g \leq 0.7$). 
However, as explained, the corresponding energies are not guaranteed to decrease monotonically.
Indeed, this is visible in Fig.~\ref{fig:Energies_Nl8_Np4_g0.2to0.6} under the form of small oscillations in the relative error starting from 15-20 iterations.
Also for this reason, in this scheme the iterative process stops earlier than in the others, making it less interesting if one is able to reach a large number of iterations.

Let us remark that the algorithms presented here reach an excellent accuracy on the total energy by spanning only a fraction of the full Hilbert space. 
Indeed, in the present case, taking into account normalization and time-reversal symmetry, one needs $\binom{8}{4} - 1 = 69$ real coefficients, hence $69$ iterations, to exactly expand the many-body wave function.
Figure~\ref{fig:Energies_Nl8_Np4_g0.2to0.6} shows instead that $20$ iterations are sufficient to get less than $1\%$ relative error on the correlation energy.

\begin{figure}
\begin{center}
\includegraphics[width=7.8cm]{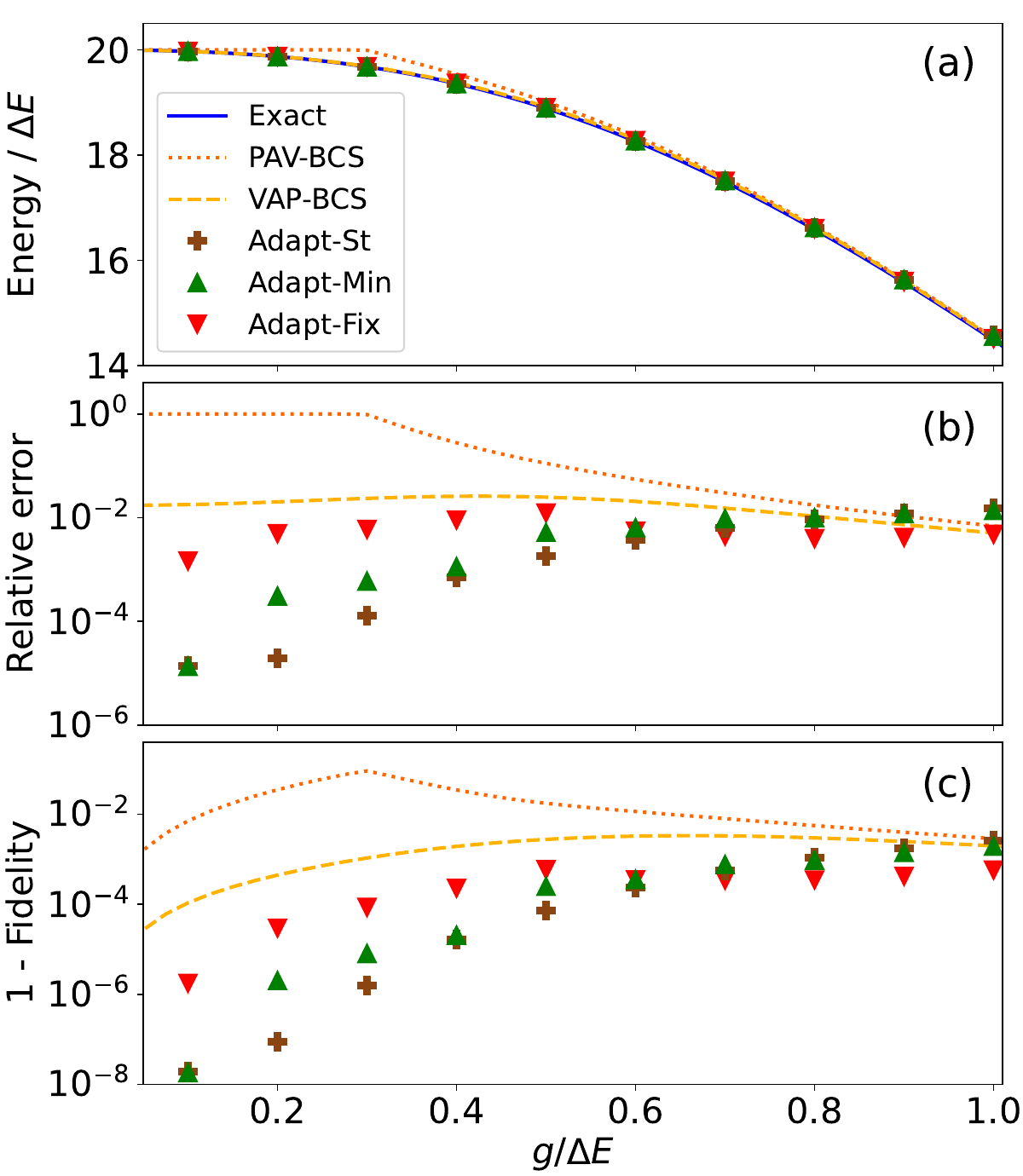}
\end{center}
\centering
\vspace{-0.4 cm}
\caption{Total ground-state energy (a), relative energy error (b), and infidelity (i.e., $1 -$ the fidelity) (c) as a function of the coupling strength $g/\Delta E$ for the different ansatzes considered in this work. As in Fig.~\ref{fig:Energies_Nl8_Np4_g0.2to0.6}, results are shown for $N=8$ particles in $D=8$ doubly-degenerated levels (half filling).
ADAPT-VQE values correspond to results after $20$ iterations.}
\label{fig:EandFidelity}
\end{figure}

In Fig.~\ref{fig:EandFidelity}, we report ground-state energies, relative errors, and fidelities obtained after 20 iterations for the same system with $g$ ranging from $0.1$ to $1.0$.
The accuracy of the adaptative methods is also compared with energies and fidelities obtained using BCS-based symmetry-breaking/symmetry-restoration ansatzes. 
Focusing on the latter, one notices that the VAP, as expected, always outperforms the PAV.
In particular, the VAP succeeds in smoothly interpolating between the normal ($g < g_c$) and superfluid ($g > g_c$) phases (in the present example $g_c \simeq 0.298$), while the PAV collapses to the Hartree-Fock limit around and below $g_c$.
In general, all methods except PAV are able to provide a good approximation to the energy for all values of the coupling strength.

The ADAPT-Fix appears to be the best approach in the large-$g$ limit, among the three ADAPT ones tested here.
However, the predicted energy, although still satisfactory, slightly degrades as $g$ decreases when compared to the other methods.
In the weak-coupling limit, it eventually becomes comparable to the VAP result. 
The two other adaptative methods, ADAPT-St and ADAPT-Min, outperform the other techniques in this regime. 
The present success of the ADAPT methods confirms the findings of Ref.~\cite{Zha24} and makes them a promising alternative to symmetry-breaking/restoration approaches for quantum computations of strongly correlated systems.
It should be noted that the VAP approach is still rather competitive, taking into account that only $D$ parameters are optimized to minimize the energy, while in adaptative methods the number of parameters equals the number of iterations. 

In the following, we use the ADAPT-VQE ansatzes obtained after 20 iterations to compute one-body Green's functions for the systems of interest, as described in Sec.~\ref{sec:strategy}.
To simplify the discussion, we will focus on the ADAPT-Min approach, which seems the best approach for $g/ \Delta E \ll 1$, i.e., for the regime of interest of nuclear physics applications.
Results are compared with those of PAV and VAP, standardly used in nuclear physics.

\subsection{Green's functions on quantum computers}
\label{sec:GFcalculation}

\subsubsection{Scaling of QPU computations}
\label{sec:scalingQPUcomp}

Starting from the ground-state approximation $|\widetilde\Psi_0^{N}\rangle$, one needs to evaluate the $4 \Omega^2$ expectation values introduced in Eqs.~\eqref{eq:overlap-H}.
As a consequence of our choice $\{A^{+}_{\alpha}\}_{\alpha=1,...,\Omega} = \{a^{\dag}_{i}\}_{i=1,\bar{1},...,D,\bar{D}}$ (and similarly for $\{A^{-}_{\alpha}\}$), $\Omega = 2D$ so one has, \textit{a priori}, $16 D^2$ expectation values to compute on the QPU. However, due to the fact that our test Hamiltonian~\eqref{eq:Hpairing} has a block-diagonal structure and $|\widetilde{\Psi}_0^N \rangle$ contains no unpaired particle, ${\cal O}^{N + 1}_{\beta \alpha} = \langle \widetilde{\Psi}_0^N | a_{\beta} a^{\dag}_{\alpha} | \widetilde{\Psi}_0^N \rangle = 0$ and 
${\cal H}^{N + 1}_{\beta \alpha} = \langle \widetilde{\Psi}_0^N | a_{\beta} H a^{\dag}_{\alpha} | \widetilde{\Psi}_0^N \rangle = 0$
for $\alpha\neq\beta$ (and similarly in the subspace with $N-1$ particles).
This implies that one needs to evaluate only $4 \Omega$ expectation values instead of $4 \Omega^2$. In addition, because of the symmetry between barred and non-barred levels in the test model in Eq.~\eqref{eq:Hpairing}, one can show that ${\cal H}^{N\pm 1}_{kk} = {\cal H}^{N\pm 1}_{\bar{k}\bar{k}}$ and ${\cal O}^{N\pm 1}_{kk} = {\cal O}^{N\pm 1}_{\bar{k}\bar{k}}$, which divide the number of required evaluations by a factor two, leading ultimately to $2 \Omega=4D$ 
expectation values to evaluate. While this reduces the numerical effort required for the results presented below, one should keep in mind that for a more general model (e.g. a realistic nuclear system), one would have to evaluate the $4 \Omega^2$ expectation values of Eqs.~\eqref{eq:overlap-H}.

%which leads to $16D^2$ expectation values to be computed on the QPU.
%For the present pairing model, however, this number can be reduced to $4D$ thanks to the symmetries of the problem.
%\vittorio{Perhaps we should specify again which symmetries?} \denis{@Samuel: I would have say "particle number symmetry" and "seniority", is there any other? } \samuel{I don't think so. It is because ${\cal O}^{N\pm 1}_{\beta \alpha} = \langle \widetilde{\Psi}_0^N | A^\mp_{\beta} A^\pm_{\alpha} | \widetilde{\Psi}_0^N \rangle = 0$ for $\alpha\neq\beta$ due to the choice of $\{A^\pm_{\alpha}\}_{\alpha}$, and ${\cal H}^{N\pm 1}_{\beta \alpha} = \langle \widetilde{\Psi}_0^N | A^\mp_{\beta} H A^\pm_{\alpha} | \widetilde{\Psi}_0^N \rangle = 0$ for $\alpha\neq\beta$ due to the choice of $\{A^\pm_{\alpha}\}_{\alpha}$ and the Hamiltonian. This implies one needs to evaluate only $4 \Omega$ expectation values instead of $4 \Omega^2$. And because of the symmetry between barred and non-barred levels, ${\cal H}^{N\pm 1}_{kk} = {\cal H}^{N\pm 1}_{\bar{k}\bar{k}}$ and ${\cal O}^{N\pm 1}_{kk} = {\cal O}^{N\pm 1}_{\bar{k}\bar{k}}$, which reduces again the number of required evaluations by a factor two.}{}
By means of the JW mapping, expressions~\eqref{eq:overlap-H} are decomposed into expectation values of Pauli terms.
%While overlaps ${\cal O}^{N\pm 1}_{\beta \alpha}$ decompose into a few Pauli terms, each reduced Hamiltonian element ${\cal H}^{N\pm 1}_{\beta \alpha}$ generates $O(D^2)$ Pauli terms.  \textcolor{blue}{While each overlap ${\cal O}^{N\pm 1}_{\alpha \alpha}$ requires one Pauli term evaluation, each reduced Hamiltonian element ${\cal H}^{N\pm 1}_{\alpha \alpha}$ generates $4(2D^2 - 5D + 4) -1$ Pauli terms, within our choice $\{A^+_{\alpha}\}_{\alpha} = \{a^{\dag}_{\alpha}\}_{\alpha}, \{A^-_{\alpha}\}_{\alpha} = \{a_{\alpha}\}_{\alpha}$, in expressions~\eqref{eq:overlap-H} and our test Hamiltonian~(\ref{eq:Hpairing}). As a result, one needs to evaluate a total of $8D(2D^2 - 5D + 4)$ Pauli terms to obtain all the needed components of the one-body Green's function. But many are similar... $D(8D^2 - 23D + (D-1)/2)$ different Pauli terms, if one supposes that the symmetry between barred and non barred indices is not broken. Or better to put it in appendix ?}
As a result, one needs to evaluate $O(D^3)$ Pauli terms to obtain all the needed components of the one-body Green's function. Taking advantage of the fact that commuting Pauli terms can be evaluated together from the same circuit, the number of required circuit measurements is lower ($O(D)$ in our specific case, as explained in appendix~\ref{App:PauliGrouping}).
While simulations presented in the following have been performed in the limit of an infinite number of measurements, numerical estimates for the number of measurements required in practice to achieve a given precision on the Green's functions are discussed in appendix~\ref{App:ErrorEstimates}.
Finally, let us note that the cost of the classical part of the calculation, i.e., the solution of the generalized eigenvalue problem~\eqref{eq:geneigen}, eventually providing the building blocks of the Lehmann representation~\eqref{eq:gfapproxlehmann}, is negligible compared to the QPU computation.

\subsubsection{Odd systems}
\label{sec:odd}

An interesting byproduct of the use of the QSE approach is that one has access to approximate eigenenergies of the neighboring odd systems with $N\pm1$ particles.
In this section we analyze the quality of the description of these odd systems for the different approaches considered in this study.
\begin{figure*}
\includegraphics[width=0.95\linewidth]{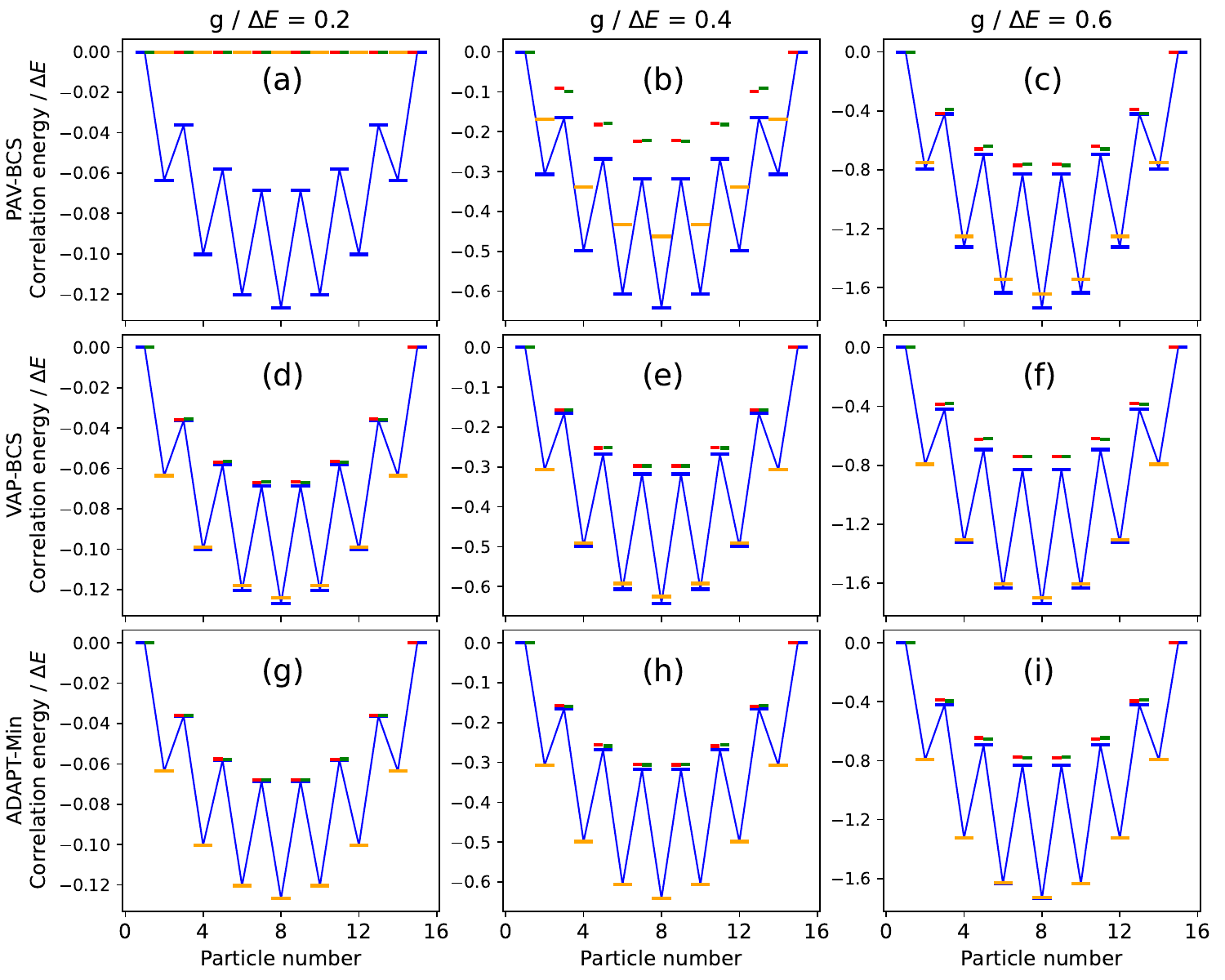}
\centering
\caption{Odd-even staggering of correlation energies obtained for $N=1$ to $N=15$ particles in a system of $D=8$ doubly degenerated levels. Different columns display results for increasing coupling strengths, namely $g=0.2$, $g=0.4$, and $g=0.6$. Exact results obtained by FCI are shown systematically in blue and compared with the approximated ones obtained from the ground state ansatzes PAV-BCS, VAP-BCS, and ADAPT-Min (from top to bottom) calculated only for even systems. Approximate energies of even systems (in orange) are directly obtained from these ground state computations, while the odd systems ones are built using the method described in section~\ref{sec:strategy} from the surrounding even systems (in red from the even system with one particle less, and in green from the one with one particle more).
}
\label{fig:OddEvenStaggering_9figs}
\end{figure*}
Figure~\ref{fig:OddEvenStaggering_9figs} shows the approximate correlation energies obtained for odd and even systems in the case of $D=8$ doubly degenerated levels and for three different values of the interaction strength.
For even systems, the energies are the direct outcome of the PAV-BCS (top panels), VAP-BCS (middle panels) and ADAPT-Min (bottom panels) methods.
Associated accuracies (and fidelities of the corresponding ground-state wave functions) are those reported in Fig.~\ref{fig:EandFidelity}.
For odd systems, energies are computed via the eigenvalue problem~\eqref{eq:geneigen}, with $|\widetilde\Psi_0^{N}\rangle$ coming consistently from the three methods.

Below the normal-to-superfluid transition, e.g., for $g=0.2$, PAV-BCS identifies with the Hartree-Fock solution and is unable to account for the odd-even staggering observed in the correlation energy.
As expected, the method nevertheless improves when moving towards the strong-coupling regime $g \gg g_c$, yielding reasonable results already at $g=0.6$.
Both VAP-BCS and ADAPT-Min provide an overall good description of odd and even systems for all interaction strengths.
A closer look reveals that ADAPT-Min is systematically more accurate, as will also be evident in the calculation of Green's functions discussed in the following.
Note also that computing ground-state energies of odd systems from below (i.e., from the even system with one less particle) or from above yields very similar results, which reinforces the consistency of the present method.

The good performance of this configuration-mixing approach suggests that it might also be applied in classical computations for more traditional nuclear structure methods, e.g., based on energy density functionals.
In that case, by starting from a PAV- or a VAP-BCS trial wave-function in the even system, one could access states in odd nuclei, which are typically harder to tackle with standard techniques. 
Other interesting aspects are that (i) because of the subspace diagonalization, the method not only gives the ground state but also excited states, and (ii) the complexity of the set of operators $\{A^{\pm}_\alpha\}_\alpha$ can be increased at will either by including 2-particle--1-hole ($2p$-$1h$) and $2h$-$1p$, ... excitations or by changing the nature of QSE operators.

\begin{figure*}
\includegraphics[width=0.95\linewidth]{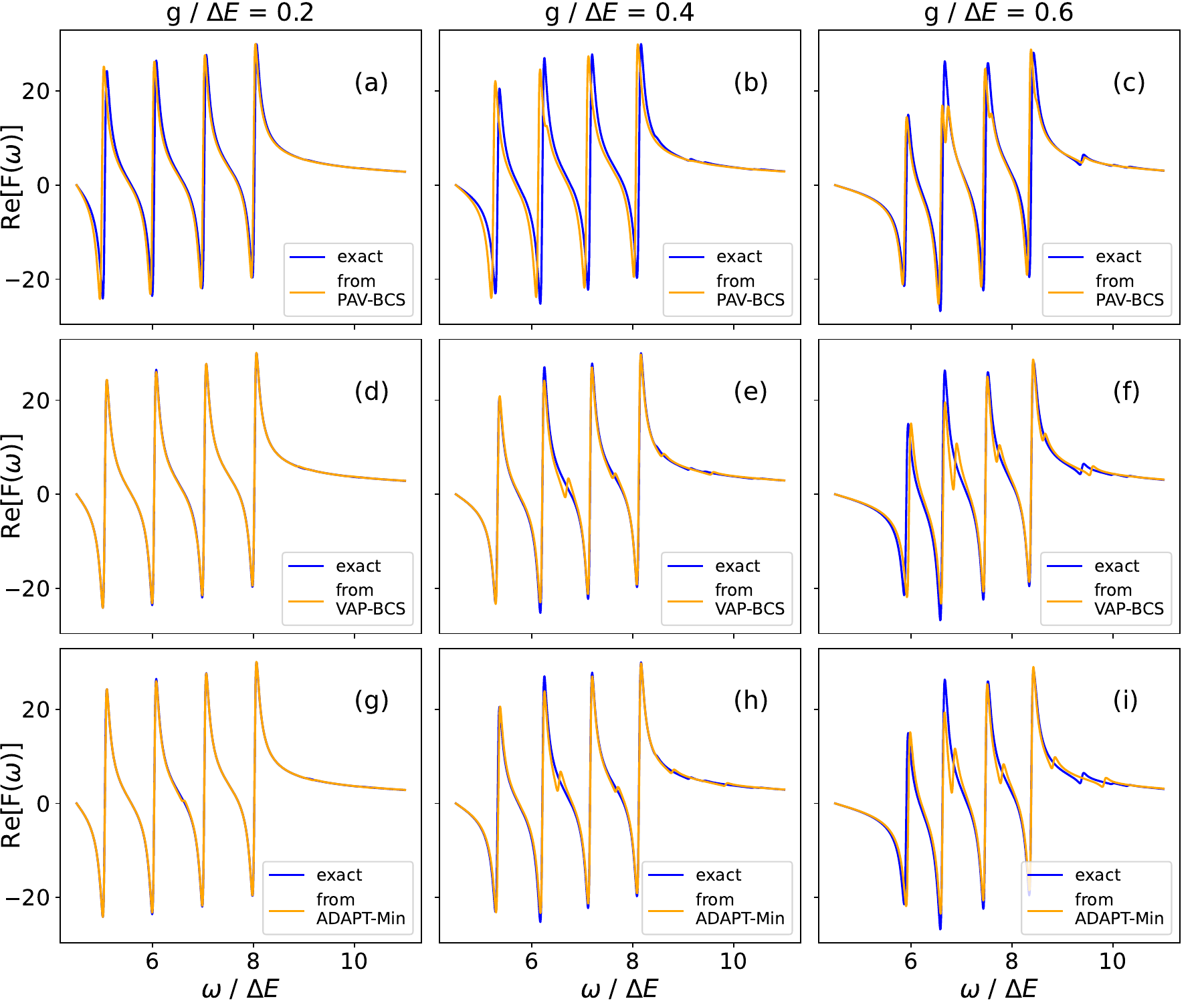}
\caption{Real part of the function $F(\omega)$ defined in Eq.~(\ref{eq:traceG}), where the regularizing parameter $\eta$ has been set to $0.04$. Taking advantage of the symmetry of $F(\omega)$ (see Fig.~\ref{fig:Exact_GFT}), for simplicity we focus here on the forward (i.e., one-particle addition) part.
Exact results for $N=8$ particles in $D = 8$ doubly-degenerated levels with coupling $g\in\{0.2,0.4,0.6\}$ (from left to right) are shown in blue and compared with approximate ones obtained with the PAV-BCS [panels (a), (b), (c)], VAP-BCS [panels (d), (e), (f)], and ADAPT-Min [panels (g), (h), (i)] techniques.}
\label{fig:GreenFunc_9plots_Nl8_Np4}
\end{figure*}

\subsubsection{Results for Green's functions}

We now turn to the main objective of the article, i.e., the evaluation of one-body Green's functions for small superfluid systems following the strategy highlighted in Sec.~\ref{sec:strategy}. 
Results based on the PAV, VAP, and ADAPT-Min approximations of the ground state are shown in Fig.~\ref{fig:GreenFunc_9plots_Nl8_Np4} and compared with exact calculations.
The same parameters appearing in Fig.~\ref{fig:OddEvenStaggering_9figs} are used here.
For simplicity, we focus on the real part of the function $\omega \mapsto F(\omega)$ defined in Eq.~(\ref{eq:traceG}).

In all cases, a fairly good reproduction of the exact Green's function is obtained independently of the coupling strength. 
Even in the PAV case, the Green's function appears rather close to the exact solution.
At weak coupling, this is mainly due to the fact that the total energy is largely dominated by the Hartree-Fock component, and even a significant error on the correlation energies does not degrade substantially the Green's function estimate.
In the intermediate regime, even if absolute correlation energies are not well reproduced by the PAV (see panel (b) of Fig.~\ref{fig:OddEvenStaggering_9figs}), such an error cancels out to a good extent when considering the energy differences that enter the denominator of Eq.~\eqref{eq:gfapproxlehmann}.
Eventually, PAV generally improves compared to the other techniques as the coupling strength increases.
While VAP and ADAPT-Min are practically indistinguishable from the exact results at small coupling, some small discrepancies become visible in the superfluid regime.

\begin{table*}[]
    \centering
    \setlength{\tabcolsep}{8pt}
    \begin{tabular}{|c|c c c|c c c|c c c|}
        \hline
        ~ & \multicolumn{3}{|c|}{PAV-BCS} & \multicolumn{3}{|c|}{VAP-BCS} & \multicolumn{3}{|c|}{ADAPT-Min} \\
        ~ & $g=0.2$ & $g=0.4$ & $g=0.6$ & $g=0.2$ & $g=0.4$ & $g=0.6$ & $g=0.2$ & $g=0.4$ & $g=0.6$ \\
        \hline
        $~~M_1~~$ & 2.5 & 5.4 & 3.0 & 0.06 & 0.3 & 0.3 & 0.002 & 0.002 & 0.3 \\
        $~~M_2~~$ & 0.8 & 2.4 & 2.5 & 0.1 & 0.5 & 1.0 & 0.08 & 0.4 & 0.9 \\
        $~~M_3~~$ & 2.9 & 6.9 & 5.3 & 0.3 & 1.2 & 2.0 & 0.2 & 0.7 & 1.8 \\
        $~~M_4~~$ & 2.7 & 7.9 & 8.1 & 0.5 & 2.3 & 4.3 & 0.4 & 1.8 & 3.9 \\
        \hline
    \end{tabular}
    \caption{Relative error (in $\%$) on the first four moments of the spectral strength distribution, Eq.~\eqref{eq:SSD}, for the PAV, VAP, and ADAPT-Min ground-state ansatzes. Values for the three coupling strengths $g = 0.2, 0.4, 0.6 $ are reported.}
    \label{SSD_table}
\end{table*}

In order to gauge the performance of the various methods in a more quantitative way, we consider the first few moments of the spectral strength distribution, defined as (see e.g. Ref.~\cite{Dug15})
\begin{equation}
    M_p \equiv \sum\limits_k S\!F_k^+ (E_k^{N+1} - E^N_0)^p 
    + \sum\limits_{k'} S\!F_{k'}^- (E_0^N - E_{k'}^{N-1})^p \,  ,
    \label{eq:SSD}
\end{equation}
where $S\!F_k^+$ and $S\!F_{k'}^-$ are the spectroscopic factors 
\begin{subequations}
\begin{align}
    S\!F_k^+ &\equiv  \sum\limits_{i} |\langle\Psi_k^{N+1}|a^\dagger_i|\Psi_0^N\rangle|^2 \, , \\
    S\!F_{k'}^- &\equiv  \sum\limits_{i} |\langle\Psi_{k'}^{N-1}|a_i|\Psi_0^N\rangle|^2 \, .
\end{align}
\end{subequations}
Relative errors with respect to exact moments (with $p=1,2,3,4$) for the different approaches are reported in Tab.~\ref{SSD_table}.
A clear hierarchy between the three methods emerges, with the PAV-BCS beginning considerably worse than the other two.
While VAP-BCS and ADAPT-Min perform generally very well, with errors of at most a few percent for the stronger coupling, the latter approach does yield lower errors across all considered cases.

\section{Conclusions}
\label{sec:conclusions}

We presented an end-to-end strategy based on a hybrid quantum-classical algorithm to obtain Green's functions 
for small superfluid systems. 
The algorithm is based on (i) obtaining an accurate approximation of the $N$-body ground state using a
hybrid quantum-classical variational method, (ii) computing approximate sets of ground- and excited states of the neighboring $(N\pm1)$-body systems using QSE tools, and (iii) making use of the Lehmann representation to construct the Green's function. 
While the manipulation of the variational ansatz for the $N$-body ground state and the calculation of the various expectation values needed in steps (i) and (iii) are performed on the QPU, some other tasks like parameter optimization and subspace diagonalizations are taken care of at the level of the CPU.

The strategy was validated using the pairing Hamiltonian at different interaction strengths. 
Several variational wave functions have been tested as a starting point.
These include ansatzes based either on the traditional (classical) symmetry-breaking/symmetry-restoration framework or on the quantum-specific ADAPT-VQE technique.
The latter approach has proven very efficient in providing an accurate approximation of the $N$-body ground state, with very high fidelities ($>99.9\%$) reached after 15-20 iterations, even as the system undergoes the normal-to-superfluid transition. 
The corresponding one-body Green's functions are close to the exact, with low moments differing only by at most a few percent.
While the performance of PAV-BCS is unsatisfactory, especially at weak coupling, the VAP variant yields good results, qualitatively close to the more advanced ADAPT-VQE methods in most cases.

As a byproduct, this scheme provides approximate eigenstates and energies of neighboring $(N\pm1)$-body systems.
Here, the QSE gives a rather accurate account of odd ground-state energies, especially for the ADAPT-Min method.
Such methodology might offer an interesting alternative for the description of odd systems on classical computers, e.g., using state-of-the-art methods based on Energy Density Functional theory.  

Present results indicate that the hybrid quantum-classical algorithm considered here offers a promising path for the calculation of Green's functions on quantum devices. 
While application to small systems has proven successful and preliminary calculations with a larger number of particles have encountered no specific obstacles, future work will have to assess more thoroughly the scalability of the method with the system size.
An important finding (already observed in Ref.~\cite{Zha24} for ground states) is that the symmetry-conserving ADAPT-VQE technique succeeds in providing an excellent description of the interacting system even in the presence of strong collective correlations (of pairing type in the present case).
This is encouraging in view of applying this approach, in the long term, to realistic nuclear Hamiltonians.
Let us also note that a generalization of the present scheme to higher-body (e.g., two-body) Green's functions would be straightforward, although clearly more challenging from the computational point of view.
In the future, it will be interesting to envisage applications to real quantum devices.
In this context, the key features of the Lehmann-based approach put forward here, i.e., the fact that (i) ($N\pm1$)-body systems are computed consistently from the neighboring $N$-body ground state and (ii) only energy differences enter the final expression for the Green's function, allow for error cancellations that might help mitigate the impact of noise.

In more strongly correlated regimes, or near phase transitions, multi-particle excitations and collective modes may become important. In such cases, extending the operator pool to include higher-body operators (e.g., $2p$–$1h$, $2h$–$1p$, or collective excitations) may be required to capture the full spectral strength. Notably, we applied the technique to the Fermi-Hubbard model, which exhibits a phase transition associated with specific spin orientations. This case is slightly more involved than the pairing Hamiltonian, since the canonical basis differs from the original single-particle basis, resulting in greater fragmentation. We observed in that case that a satisfactory reproduction of the exact Green’s function can be achieved in two ways: either (i) one fully optimizes the initial state to reproduce the exact ground state with $N$ particles and restricts $N\pm 1$ subspaces to $1p$ and $1h$ excitations, or (ii) one computes variationally an approximate $N$-body ground state and then includes $2p$–$1h$, $2h$–$1p$ or higher-order excitations. In general, the effectiveness of the method relies on an accurate description of the $N$-body ground state and the richness of the QSE operator set.

The work presented here can be regarded as a proof of principle in view of future applications that are relevant to nuclear physics.
One can anticipate that the extension of the current approach to realistic nuclear Hamiltonians will introduce several additional challenges.
First, as discussed in appendix~\ref{app:ExpectValuesEval}, the symmetries of the pairing Hamiltonian induce a dramatic reduction in the measurements required for reconstructing the overlaps and Hamiltonian kernels. For a general Hamiltonian, a significant increase of the number of measurements is to be expected.
Second, realistic interactions generally produce stronger fragmentation of the spectral strength as well as collective excitations (e.g., deformation, pairing vibrations, clustering phenomena).
As discussed above, this may require larger operator pools and enlarged QSE subspaces.
In particular, in future applications one may envisage the use of QSE operators that effectively account for such collective modes as an alternative to independent multi-particle multi-hole excitations.

In general, we expect that the growth of the QSE subspace and the measurement overhead will constitute the primary scalability limitations.
The use of more collective operator pools or symmetry restoration will be key to maintaining accuracy for future applications in realistic systems.

\section*{Acknowledgments }

DL thanks S. Baid and J. M. Arias Carrasco for the discussions on the ADAPT-VQE at the early stage of this work. This project has received financial support from the CNRS through the AIQI-IN2P3 project.
JZ is funded by the joint doctoral programme of Universit\'e Paris-Saclay and the Chinese Scholarship Council.
This work is part of 
HQI initiative (\href{www.hqi.fr}{www.hqi.fr}) and is supported by France 2030 under the French 
National Research Agency award number ``ANR-22-PNQC-0002''.
Calculations were performed using HPC resources from GENCI-TGCC (Contracts No. A0170513012 and AD010615269). 

\appendix 

%\section{Jordan-Wigner mapping with pair operators}
\section{Jordan-Wigner mappings: particle versus pair encoding}
\label{app:CoherenceJWTpairs}

\subsection{Particle-to-qubit encoding of Hamiltonian~(\ref{eq:Hpairing})}

Using the Jordan-Wigner transformation $JW[.]$ as defined in Eqs.~(\ref{EqJWT}), Hamiltonian~(\ref{eq:Hpairing}) can be developed into Pauli operators as:
\begin{align}
    JW[H] &= JW\Big[\sum\limits_{k=1}^{D} \varepsilon_k (a_k^{\dag}a_k + a_{\bar{k}}^{\dag}a_{\bar{k}}) 
    - g\sum\limits_{k\neq l} a_k^{\dag} a_{\bar{k}}^{\dag} a_{\bar{l}}a_l\Big] ,\nonumber \\
    &= \sum\limits_{k=1}^{D} \varepsilon_k (\mathds{1} - \frac{1}{2}Z_k - \frac{1}{2}Z_{\bar{k}}) 
    - g\Gamma.
\end{align}
where
\begin{eqnarray}
    \Gamma &=& \!\sum\limits_{k \neq l} \! \sigma^-_k \!\sigma^-_{\bar{k}} \!\sigma^+_{\bar{l}} \!\sigma^+_l = \frac{1}{8} \!\sum\limits_{k < l}
    \!\Big[ X_k X_{\bar{k}} X_{\bar{l}} X_l + Y_k  Y_{\bar{k}}Y_{\bar{l}} Y_l \nonumber \\[-8pt]
    && \hspace*{35mm} -~ X_k X_{\bar{k}} Y_{\bar{l}} Y_l - Y_k Y_{\bar{k}} X_{\bar{l}} X_l \nonumber \\[3pt]
    && \hspace*{35mm} +~ X_k Y_{\bar{k}} X_{\bar{l}} Y_l  + Y_k X_{\bar{k}} Y_{\bar{l}} X_l \nonumber \\
    && \hspace*{35mm} +~ X_k Y_{\bar{k}} Y_{\bar{l}} X_l  + Y_k X_{\bar{k}} X_{\bar{l}} Y_l  \Big]. \nonumber
\end{eqnarray}
This scheme assumes that each single-particle state is encoded on a qubit, and thus requires $2D$ qubits for $D$ doubly-degenerated single-particle levels. 

\subsection{From particle-to-qubit to pair-to-qubit encoding}
%\subsection{Coherence between the two encodings}

As mentioned in the main text, for even systems, one can reduce the number of qubits by a factor two, by directly encoding each pair occupation/non-occupation on a single qubit.
Here we show how this simplification can be justified from the particle-to-qubit encoding.

\color{black}

The pair operators $P_k^{\dag} = a_k^{\dag} a_{\bar{k}}^{\dag}$ and $P_k = a_{\bar{k}} a_k$ can be mapped into Pauli operators using the original Jordan-Wigner mapping JW[.] for individual particles defined in Eqs.~(\ref{EqJWT}) as
\begin{align}
    {\rm JW}[P_k^{\dag}] &= {\rm JW}[a_k^{\dag}] {\rm JW}[a_{\bar{k}}^{\dag}] \nonumber\\
    &= \frac{1}{4}\Big[X_k X_{\bar{k}} - Y_k Y_{\bar{k}} - i(X_k Y_{\bar{k}} + Y_k X_{\bar{k}})\Big]
    \label{eq:JWpairs}
\end{align}
up to a global sign that depends on the order of the indices (i.e., whether $k<\bar{k}$ or $\bar{k}<k$). 
%\textcolor{blue}{Using it, Hamiltonian~(\ref{eq:Hpairing_pairs}) can be developed into Pauli terms as :
%\begin{align}
%    JW[H] &= \sum\limits_{k=1}^{D} 2\varepsilon_k JW[P_k^{\dag}P_k] - %g\sum\limits_{k\neq l} JW[P_k^{\dag} P_l] \nonumber\\
%    &= \sum\limits_{k=1}^{D} \frac{\varepsilon_k}{2} (\mathds{1} - Z_k - %Z_{\bar{k}} + Z_k Z_{\bar{k}}) - g\sum\limits_{k\neq l} JW[P_k^{\dag} P_l]
%\end{align}
%}

The matrix representation of the operators on the second line of Eq.~\eqref{eq:JWpairs} in the two-qubits basis $\{ |00\rangle_{k\bar{k}}, |01\rangle_{k\bar{k}}, |10\rangle_{k\bar{k}}, |11\rangle_{k\bar{k}} \}$ reads as
\begin{subequations}
\begin{align}
    &X_k X_{\bar{k}} = \begin{pmatrix} 0 & \hspace{4pt} 1 \\ 1 & \hspace{4pt} 0 \end{pmatrix}_{\!k} \!\otimes \begin{pmatrix} 0 & \hspace{4pt} 1 \\ 1 & \hspace{4pt} 0 \end{pmatrix}_{\!\bar{k}} = \begin{psmallmatrix} \vspace{2pt} 0 & \hspace{2pt} 0 & \hspace{2pt} 0 & \hspace{2pt} 1 \\ \vspace{2pt} 0 & \hspace{2pt} 0 & \hspace{2pt} 1 & \hspace{2pt} 0 \\ \vspace{2pt} 0 & \hspace{2pt} 1 & \hspace{2pt} 0 & \hspace{2pt} 0 \\ 1 & \hspace{2pt} 0 & \hspace{2pt} 0 & \hspace{2pt} 0 \end{psmallmatrix}_{\!k\bar{k}} \, ,  \\
    &X_k Y_{\bar{k}} = \begin{pmatrix} 0 & \hspace{4pt} 1 \\ 1 & \hspace{4pt} 0 \end{pmatrix}_{\!k} \!\otimes \begin{pmatrix} 0 & -i \\ i & 0 \end{pmatrix}_{\!\bar{k}} = \begin{psmallmatrix}  \vspace{2pt} 0 & 0 & 0 & -i \\ \vspace{2pt} 0 & 0 & i & 0 \\ \vspace{2pt} 0 & -i & 0 & 0 \\ i & 0 & 0 & 0 \end{psmallmatrix}_{\!k\bar{k}} \, ,  \\
    &Y_k X_{\bar{k}} = \begin{pmatrix} 0 & -i \\ i & 0 \end{pmatrix}_{\!k} \!\otimes \begin{pmatrix} 0 & \hspace{4pt} 1 \\ 1 & \hspace{4pt} 0 \end{pmatrix}_{\!\bar{k}} = \begin{psmallmatrix}  \vspace{2pt} 0 & \hspace{2pt} 0 & 0 & -i \\ \vspace{2pt} 0 & \hspace{2pt} 0 & -i & 0 \\ \vspace{2pt} 0 & \hspace{2pt} i & 0 & 0 \\ i & \hspace{2pt} 0 & 0 & 0 \end{psmallmatrix}_{\!k\bar{k}} \, ,  \\
    &Y_k Y_{\bar{k}} = \begin{pmatrix} 0 & -i \\ i & 0 \end{pmatrix}_{\!k} \!\otimes \begin{pmatrix} 0 & -i \\ i & 0 \end{pmatrix}_{\!\bar{k}} = \begin{psmallmatrix}  \vspace{2pt} 0 & 0 & \hspace{2pt} 0 & -1 \\ \vspace{2pt} 0 & 0 & \hspace{2pt} 1 & 0 \\ \vspace{2pt} 0 & 1 & \hspace{2pt} 0 & 0 \\ -1 & 0 & \hspace{2pt} 0 & 0 \end{psmallmatrix}_{\!k\bar{k}} \, .
\end{align}
\end{subequations}
Let us now define 
\begin{subequations}
\begin{eqnarray}
\widetilde{X} &\equiv \frac{1}{2}(X_k X_{\bar{k}} - Y_k Y_{\bar{k}})  = \begin{psmallmatrix} \vspace{2pt} 0 & \hspace{3pt} 0 & \hspace{3pt} 0 & \hspace{3pt} 1 \\ \vspace{2pt} 0 & \hspace{3pt} 0 & \hspace{3pt}0 & \hspace{3pt} 0 \\ \vspace{2pt} 0 & \hspace{3pt} 0 & \hspace{3pt} 0 & \hspace{3pt} 0 \\ 1 & \hspace{3pt} 0 & \hspace{3pt} 0 & \hspace{3pt} 0 \end{psmallmatrix}_{\!k\bar{k}} \!\! \, , \\
\widetilde{Y} &\equiv \frac{1}{2}(X_k Y_{\bar{k}} + Y_k X_{\bar{k}}) = \begin{psmallmatrix} \vspace{2pt} 0 & \hspace{3pt} 0 & \hspace{3pt} 0 & -i \\ \vspace{2pt} 0 & \hspace{3pt} 0 & \hspace{3pt}0 & \hspace{1pt} 0 \\ \vspace{2pt} 0 & \hspace{3pt} 0 & \hspace{3pt} 0 & \hspace{1pt} 0 \\ i & \hspace{3pt} 0 & \hspace{3pt} 0 & \hspace{1pt} 0 \end{psmallmatrix}_{\!k\bar{k}}\!\! \, .
\end{eqnarray}
\end{subequations}
These operators can be seen as effectively acting only on span($|00\rangle_{k\bar{k}}, |11\rangle_{k\bar{k}}$), i.e., on the seniority-zero subspace considered in the present work.
Moreover, one realizes that in this subspace, which can be naturally mapped to a single qubit, $\widetilde{X}$ and $\widetilde{Y}$ read precisely as $X$ and $Y$ Pauli matrices. 
It follows that $P_k$ and $P_k^{\dag}$ can be utilized when considering pairs, exactly like $a_k$ and $a_k^{\dag}$ are employed when considering particles, and one can similarly apply Jordan-Wigner transformations for both. Below we show how this works for the pair-encoding case.

\subsection{Pair-to-qubit encoding of Hamiltonian~(\ref{eq:Hpairing})}

In the pair-to qubit encoding, one directly encodes the creation/annihilation operator on qubits using the transformation
\begin{eqnarray}
    P^\dagger_k \rightarrow \frac{1}{2} \left( X_k - i Y_k\right), ~~
    P_k \rightarrow \frac{1}{2} \left( X_k + i Y_k\right) \, .
    \label{eq:pairqubit}
\end{eqnarray}
Provided that we restrict the system to the Hilbert subspace of seniority $0$, one can transform the one-body part of the Hamiltonian as
\begin{eqnarray}
    \varepsilon_k (a_k^{\dag}a_k + a_{\bar{k}}^{\dag}a_{\bar{k}})  ~ \sim ~ 2 \varepsilon_k P^\dagger_k P_k \, , 
\end{eqnarray}
which leads to the expression~\eqref{eq:Hpairing_pairs}. Noteworthy, the relation between these two operators only holds in the seniority $0$ subspace, and could not be used otherwise.

Using now the pair-to-qubit mapping of Eq.~\eqref{eq:pairqubit}, one obtains
\begin{eqnarray}
    H_P &=&  \sum\limits_{k=1}^{D} \varepsilon_k (\mathds{1} - Z_k) - \frac{g}{2}\sum\limits_{k<l} (X_k X_l + Y_k Y_l) \, , 
\end{eqnarray}
which corresponds to the one used in Refs. \cite{Kha21,Rui22,Zha24}, or in more recent publications \cite{yoshida2024,Yos25,Cos25} in the limit of constant pairing 
strength.

%\color{blue}
%To compare particle-to-qubit encoding and pair-to-qubit encoding, one can write the Hamiltonian~(\ref{eq:Hpairing}) as a combination of Pauli terms within the two schemes. With particle-to-qubit encoding :
%\begin{align}
%    JW[H] &= JW\Big[\sum\limits_{k=1}^{D} \varepsilon_k (a_k^{\dag}a_k + a_{\bar{k}}^{\dag}a_{\bar{k}}) 
%    - g\sum\limits_{k\neq l} a_k^{\dag} a_{\bar{k}}^{\dag} a_{\bar{l}}a_l\Big] \nonumber \\
%    &= \sum\limits_{k=1}^{D} \frac{\varepsilon_k}{2} (2\mathds{1} - Z_k - Z_{\bar{k}}) 
%    - g\sum\limits_{k\neq l} JW[a_k^{\dag} a_{\bar{k}}^{\dag} a_{\bar{l}}a_l]
%\end{align}
%while with pair-to-qubit encoding :
%\begin{align}
%    JW[H] &= JW\Big[\sum\limits_{k=1}^{D} 2\varepsilon_k P_k^{\dag}P_k - g\sum\limits_{k\neq l} P_k^{\dag} P_l\Big] \nonumber \\
%    &= \sum\limits_{k=1}^{D} \varepsilon_k (\mathds{1} - Z_k) - g\sum\limits_{k\neq l} (X_k X_l + Y_k Y_l + i(X_k Y_l - Y_k X_l))
%\end{align}

\color{black}

\section{Justification of Eq.~(\ref{EnergySineForm}) and determination of its parameters}

\subsection{Justification}
\label{app:DemoEq23}

The QEB-Pool operators $G_{ij} = \frac{1}{2}(X_i Y_j - Y_i X_j)$, for $i \neq j$, considered in this work satisfy the property $G_{ij}^3 = G_{ij}$. 
Here we show that this property is sufficient to obtain Eq.~\eqref{EnergySineForm}.

Let G be an operator satisfying $G^3 = G$. Then, for $p\in\mathbb{N}$
\begin{equation}
    G^p = \left\{
    \begin{array}{ll}
        \mathds{1} & \text{if } p=0 \\
        G & \text{if $p$ is odd} \\
        G^2 & \text{if $p$ is even} \neq 0.
    \end{array}
\right.
\end{equation}
~

This allows to write, for any angle $\theta$,
\begin{align}
    e^{i\theta G} &= \sum\limits_{p=0}^{\infty} \frac{(i\theta G)^p}{p!} \nonumber \\
    &= \mathds{1} + \sum\limits_{p=1}^{\infty} \frac{(-1)^p\theta^{2p}}{(2p)!} G^2 + i\sum\limits_{p=0}^{\infty} \frac{(-1)^p\theta^{2p+1}}{(2p+1)!} G \nonumber \\
    &= \mathds{1} + (\cos(\theta) - 1) G^2 + i\sin(\theta)G \, .
\end{align}
By inserting this expression in $E(\theta) = \langle\psi|e^{-i\theta G} \!H\! e^{i\theta G}|\psi\rangle$, developing all terms and exploiting some trigonometric identities, one can finally put it in the form of Eq.~\eqref{EnergySineForm}
\begin{equation}
E(\theta) = C_0 + C_1\cos(\theta + \gamma_1) + C_2\cos(2\theta + \gamma_2) \, .
\end{equation}

\subsection{Determination of the parameters}
\label{app:FixingParameters}

The five parameters ($C_0,C_1,C_2,\gamma_1,\gamma_2$) introduced in Eq. (\ref{EnergySineForm}) can be determined from five evaluations of $E(\theta)$. As an example, let us consider
\begin{subequations}
\begin{align}
    E(0) &= C_0 + C_1\cos(\gamma_1) + C_2\cos(\gamma_2)\: , \\
    E(\pi/2) &= C_0 - C_1\sin(\gamma_1) - C_2\cos(\gamma_2)\: , \\
    E(\pi) &= C_0 - C_1\cos(\gamma_1) + C_2\cos(\gamma_2)\: , \\
    E(3\pi/2) &= C_0 + C_1\sin(\gamma_1) - C_2\cos(\gamma_2)\: , \\
    E(\pi/4) &= C_0 + C_1\cos(\pi/4 + \gamma_1) - C_2\sin(\gamma_2)\: .
\end{align}
\end{subequations}
In this case $C_0, C_1, \gamma_1$ can be retrieved from
\begin{subequations}
\begin{align}
    &C_0 = \frac{1}{4}\Big(E(0) + E(\pi/2) + E(\pi) + E(3\pi/2)\Big)\: , \\
    &C_1\cos(\gamma_1) = \frac{1}{2}\Big(E(0) - E(\pi)\Big)\: , \\
    &C_1\sin(\gamma_1) = \frac{1}{2}\Big(E(3\pi/2) - E(\pi/2)\Big)\: .
\end{align}
\end{subequations}
Subsequently, $C_2, \gamma_2$ are deduced from
\begin{subequations}
\begin{align}
    &C_2\cos(\gamma_2) = \frac{1}{2}\Big(E(0) + E(\pi)\Big) - C \: ,\\
    &C_2\sin(\gamma_2) = C + A_1\cos(\pi/4 + \gamma_1) - E(\pi/4) \: . 
\end{align}
\end{subequations}
This allows to fully determine the function $\theta \mapsto E(\theta)$ and, more specifically, to find its global minimum. 
In the presence of noise, more evaluations can be performed to fit $E(\theta)$.

%~ % Why there is no automatic space here ?
\section{Evaluation of expectation values~\eqref{eq:overlap-H}}
\label{app:ExpectValuesEval}

Here we give additional details on how, in practice, the expectation values of the overlap and Hamiltonian kernels, given in Eqs.~(\ref{eq:overlap}) and (\ref{eq:hamil-ab}) respectively, are evaluated. 
As explained in section~\ref{sec:scalingQPUcomp}, within our test model and our choice of operators $\{A^{\pm}_\alpha\}_\alpha$, one simply needs to evaluate
\begin{subequations}
\begin{align}
    {\cal O}^{N + 1}_{ii} &= \langle \widetilde{\Psi}_0^N \!| a_{i} a^{\dag}_{i} \!| \widetilde{\Psi}_0^N \!\rangle \, , \\
    {\cal O}^{N - 1}_{ii} &= \langle \widetilde{\Psi}_0^N \!| a^{\dag}_{i} a_{i} \!| \widetilde{\Psi}_0^N \!\rangle \, ,
\end{align}
and
\begin{align}
    {\cal H}^{N + 1}_{ii} = \langle \widetilde{\Psi}_0^N \!| a_{i} H a^{\dag}_{i} \!| \widetilde{\Psi}_0^N \!\rangle \, , \\
    {\cal H}^{N - 1}_{ii} = \langle \widetilde{\Psi}_0^N \!| a^{\dag}_{i} H a_{i} \!| \widetilde{\Psi}_0^N \!\rangle \, ,
\end{align}
\end{subequations}
for $i\in\{1, ..., D\}$. 
To do so, one has to express operators $a_{i} a^{\dag}_{i}$, $a^{\dag}_{i} a_{i}$, $a_{i} H a^{\dag}_{i}$, $a^{\dag}_{i} H a_{i}$ with Pauli operators.
Then, all Pauli term expectation values are extracted from measurements on qubits encoding the state $|\widetilde{\Psi}_0^N \rangle$, after the corresponding basis rotation has been performed.
These two steps are explained in the following.

\subsection{Expansion into Pauli operators}
\label{app:PauliOpExpansion_HandO}

After expressing $a_{i} a^{\dag}_{i}$ and $a^{\dag}_{i} a_{i}$ as
\begin{subequations}
\begin{align}
    JW[a_{i} a^{\dag}_{i}] &= \frac{1}{2}(\mathds{1} + Z_i) \, , \\
    JW[a^{\dag}_{i} a_{i}] &= \frac{1}{2}(\mathds{1} - Z_i) \, ,
\end{align}
\end{subequations}
one can write the overlaps as
%where the linearity and the preservation of operator products have been used, as well as the cancellation of the two factors $\prod\limits_{p < i}\!(-Z_p)$.
\begin{subequations}
\begin{align}
    {\cal O}^{N + 1}_{ii} &= \frac{1}{2} + \frac{1}{2}\langle \widetilde{\Psi}_0^N | Z_i | \widetilde{\Psi}_0^N \rangle \, , \\
    {\cal O}^{N - 1}_{ii} &= \frac{1}{2} - \frac{1}{2}\langle \widetilde{\Psi}_0^N | Z_i | \widetilde{\Psi}_0^N \rangle \, . 
\end{align}
\end{subequations}
It follows that only one Pauli term has to be evaluated both for ${\cal O}^{N + 1}_{ii}$ and ${\cal O}^{N - 1}_{ii}$.

The derivation of the different Hamiltonian kernels ${\cal H}^{N + 1}_{ii}$ and ${\cal H}^{N - 1}_{ii}$ is less straightforward. Using the expression of Hamiltonian~(\ref{eq:Hpairing}), one has
\begin{align}
    JW[a_{i} H a^{\dag}_{i}] &= \sum\limits_{k=1}^{D} \varepsilon_k \Big(JW[a_{i}a_k^{\dag}a_ka^{\dag}_{i}] + JW[a_{i}a_{\bar{k}}^{\dag}a_{\bar{k}}a^{\dag}_{i}]\Big) \nonumber\\
    & ~~~~~~~~~~~~ - g\sum\limits_{k\neq l} JW[a_{i} a_k^{\dag} a_{\bar{k}}^{\dag} a_{\bar{l}} a_l a^{\dag}_{i}] \, , 
\end{align}
for $i\in {\cal I} \equiv\{1,\bar{1}, ~...~ , D,\bar{D}\}$. To compute the terms $JW[a_{i}a_j^{\dag}a_ja^{\dag}_{i}]$ for $j\in {\cal I}$, one has to distinguish the case $j\neq i$, for which
\begin{eqnarray}
    JW[a_{i}a_j^{\dag}a_ja^{\dag}_{i}] &= JW[a_{i}a^{\dag}_{i}]JW[a_j^{\dag}a_j] \nonumber \\ 
    &= \displaystyle \frac{1}{4}(\mathds{1} + Z_i)(\mathds{1} - Z_j) \, ,
\end{eqnarray} from the case $j = i$, for which
\begin{eqnarray}
JW[a_{i} a_j^{\dag}a_j a^{\dag}_{i}] \!=\! JW[a_{i}a^{\dag}_{i}] \!=\! \frac{1}{2}(\mathds{1} \!+\! Z_i) \, . 
\end{eqnarray}
Then, for $k,l\in \{1, ~...~ ,D\}$,
\begin{eqnarray}
    JW[a_{i} a_k^{\dag} a_{\bar{k}}^{\dag} a_{\bar{l}} a_l a^{\dag}_{i}] = JW[a_{i} a^{\dag}_{i}]JW[a_k^{\dag} a_{\bar{k}}^{\dag} a_{\bar{l}} a_l]
\end{eqnarray}
if $i\notin \{k,\bar{k},l,\bar{l}\}$, while it is $0$ otherwise.
Using the fact that $JW[a_k^{\dag} a_{\bar{k}}^{\dag} a_{\bar{l}} a_l] = \sigma^-_k \sigma^-_{\bar{k}} \sigma^+_{\bar{l}} \sigma^+_l$ with $\sigma^{\pm}_\alpha \equiv \frac{1}{2}(X_\alpha \pm iY_\alpha)$, 
%(all the $Z$'s and minus signs cancel or are absorbed with the relations $Z\sigma^{\pm} = \pm\sigma^{\pm} = - \sigma^{\pm}Z$). Finally, 
one finally obtains
\begin{align} \label{Pauliterms_H^Np1}
    JW[a_{i} H a^{\dag}_{i}] &= (\mathds{1} \!+\! Z_i) \Big( \frac{\varepsilon_i}{2}\mathds{1} + \hspace{-8pt}\sum\limits_{j\in I\backslash{\{i\}}} \hspace{-8pt} \frac{\varepsilon_j}{4} (\mathds{1} - Z_j) - \frac{g}{2}\Gamma_{\!i} \Big) \nonumber\\
    &= (\mathds{1} \!+\! Z_i) \!\!\left[\! \left( \!\frac{\varepsilon_i}{4} + \!\sum\limits_{k=1}^D \!\frac{\varepsilon_k}{2}\!\!\right)\!\!\mathds{1} - \hspace{-9pt}\sum\limits_{j\in {\cal I}\backslash{\{i\}}} \hspace{-8pt} \frac{\varepsilon_j}{4} Z_j - \frac{g}{2}\Gamma_{\!i} \!\right] \!\!,
\end{align}

\vspace{-6mm}
\noindent where
%\begin{align}
%    \Gamma_{\!i} = \!\sum\limits_{\substack{k \neq l \\ k,l \neq i}} &\sigma^-_k \!\sigma^-_{\bar{k}} \!\sigma^+_{\bar{l}} \!\sigma^+_l = \frac{1}{8} \!\!\sum\limits_{\substack{k < l \\ k,l \neq i}} \!\!\begin{pmatrix}
%        X\!X\!X\!X + Y\!Y\!Y\!Y \\
%        -~ X\!X\!Y\!Y - Y\!Y\!X\!X \\
%        +~ X\!Y\!X\!Y + Y\!X\!Y\!X \\
%        +~ X\!Y\!Y\!X + Y\!X\!X\!Y
%    \end{pmatrix}_{\!\!\!\!k \bar{k} \bar{l} l} .
%\end{align}
\begin{eqnarray}
    \Gamma_{\!i} &=& \!\sum\limits_{\substack{k \neq l \\ k,l \neq i}} \!\! \sigma^-_k \!\sigma^-_{\bar{k}} \!\sigma^+_{\bar{l}} \!\sigma^+_l = \frac{1}{8} \!\!\sum\limits_{\substack{k < l \\ k,l \neq i}} 
    \!\Big[ X_k X_{\bar{k}} X_{\bar{l}} X_l + Y_k  Y_{\bar{k}}Y_{\bar{l}}Y_l \nonumber \\[-14pt]
    && \hspace*{3.7cm} -~ X_k X_{\bar{k}} Y_{\bar{l}} Y_l - Y_k Y_{\bar{k}} X_{\bar{l}}X_l \nonumber \\[3pt]
    && \hspace*{3.7cm} +~ X_k Y_{\bar{k}} X_{\bar{l}} Y_l  + Y_k X_{\bar{k}}Y_{\bar{l}}X_l \nonumber \\
    && \hspace*{3.7cm} +~ X_kY_{\bar{k}}Y_{\bar{l}}X_l  + Y_kX_{\bar{k}}X_{\bar{l}}Y_l  \Big] \, . \nonumber
\end{eqnarray}
Note that, when developing $\sigma^-_k \sigma^-_{\bar{k}} \sigma^+_{\bar{l}} \sigma^+_l + \sigma^-_l \sigma^-_{\bar{l}} \sigma^+_{\bar{k}} \sigma^+_k$, all terms with an imaginary coefficient cancel. Similarly, one has
\begin{equation} 
    JW[a^{\dag}_{i} H a_{i}] = (\mathds{1} \!-\! Z_i) \!\!\left[\! \left( \!\textcolor{blue}{-}\frac{\varepsilon_i}{4} + \!\sum\limits_{k=1}^D \!\frac{\varepsilon_k}{2}\!\!\right)\!\!\mathds{1} - \hspace{-9pt}\sum\limits_{j\in {\cal I}\backslash{\{i\}}} \hspace{-8pt} \frac{\varepsilon_j}{4} Z_j - \frac{g}{2}\Gamma_{\!i} \!\right] \!\!.
    \label{Pauliterms_HNm1}
\end{equation}

From these expressions, we see that there are $O(D^2)$ Pauli terms to evaluate in order to get ${\cal H}^{N + 1}_{ii}$ and ${\cal H}^{N - 1}_{ii}$, mainly coming from $\Gamma_{\!i}$. This has to be done for all $i\in \{1, ..., D\}$, leading to a total of $O(D^3)$ Pauli terms needed to build the Green's function of Hamiltonian~(\ref{eq:Hpairing}).

\begin{table*}[t]
    \centering
    \includegraphics[width=1.0\linewidth]{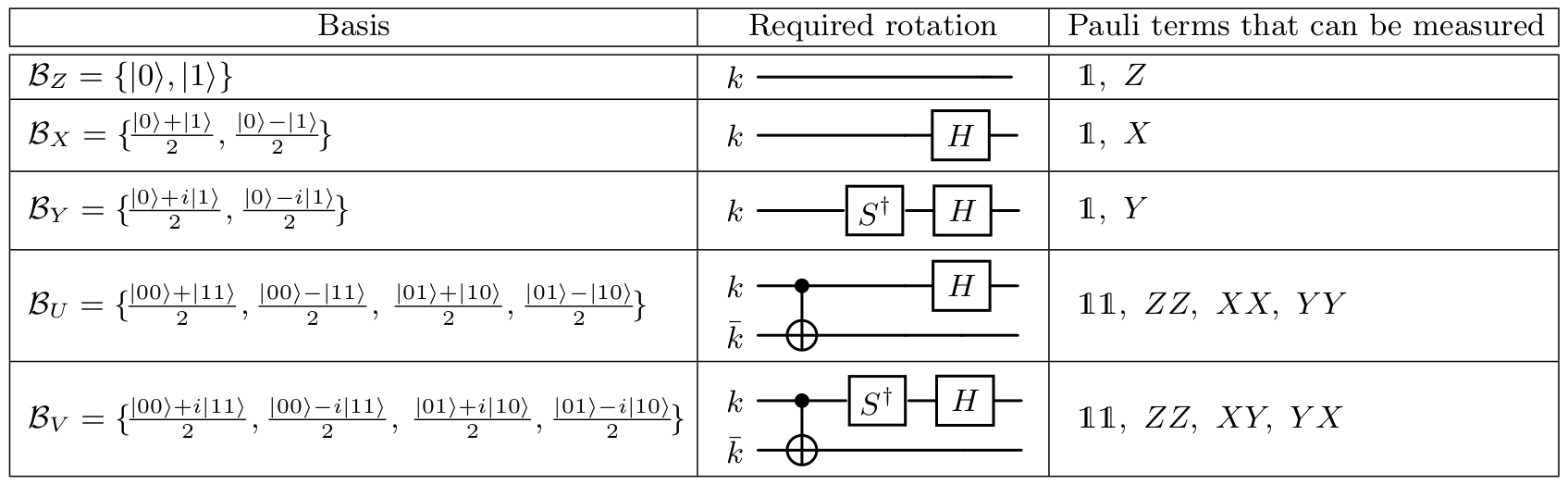}
    \caption{Different one and two qubits bases, corresponding rotations to be performed before measurement, and Pauli terms whose expectation values can be retrieved from the probabilities of the measurement outcomes.}
    \label{tab:MeasurementBases}
\end{table*}

\subsection{Evaluation of Pauli terms}
\label{App:PauliTermsEval}

The measurement of a qubit yields two possible outcomes: 0 or 1. Similarly, the measurement of a system of $N$ qubits gives a bitstring. If the same procedure (operations on qubits to prepare a state $|\Psi\rangle$ and measurement) is repeated sufficiently many times, one has access to the probabilities associated to the measurement of each bitstring. From these probabilities, one can extract the expectation value of Pauli terms containing only $\mathds{1}$'s and $Z$'s.

Indeed, if one writes such a Pauli term $Z^{a_1} \otimes ... \otimes Z^{a_N}$ where $\forall i, a_i \in \{0,1\}$ so that $Z^{a_1}$ is $\mathds{1}$ or $Z$, it can be shown that its expectation value reads as
\begin{equation}
    \langle\Psi| Z^{a_1} \otimes ... \otimes Z^{a_N} |\Psi\rangle = \hspace{-10pt} \sum\limits_{b_1...b_N \in \{0,1\}^{N}} \hspace{-10pt} (-1)^{\sum\limits_k a_kb_k} P_{|\Psi\rangle}(b_1...b_N) \, ,
    \label{Eq:PauliTermEval}
\end{equation}
where $P_{|\Psi\rangle}(b_1...b_N)$ is the probability to obtain the bitstring $b_1...b_N$ from the measurement of $|\Psi\rangle$. As a remark, all Pauli terms containing only $\mathds{1}$'s and $Z$'s can be extracted from the knowledge of probabilities $P_{|\Psi\rangle}(b_1...b_N)$ for $b_1...b_N \in \{0,1\}^{N}$.

Expectation values of other Pauli terms (including $X$'s and $Y$'s) can be obtained in the same way by adding qubit basis rotations before the measurement (namely, a Hadamard gate to measure $X$ and a $HS^{\dag}$ gate to measure $Y$, as shown in table~\ref{tab:MeasurementBases}). 
For more details, we refer to~\cite{Ayr23}. 

Let us denote $\mathcal{B}_{\sigma}$ the basis corresponding to the measurement of $\sigma\in\{X, Y, Z\}$. As before, once the basis $\mathcal{B}_{\sigma_1} \otimes ... \otimes \mathcal{B}_{\sigma_N}$ (with $\forall i\in \{1, ..., N\}, \sigma_i \in \{X,Y,Z\}$) is fixed, one can obtain all Pauli terms of the form $\sigma_1^{a_1} \otimes ... \otimes \sigma_N^{a_N}$ from the same set of measurements.
More generally, Pauli terms can be grouped together in fully commuting sets. 
For each of these sets there is a corresponding basis in which all terms of the set can be measured at once (some examples are given in table~\ref{tab:MeasurementBases}). 
This allows to reduce the total number of quantum circuit evaluations to $N_{sets} \times N_{shots}$, where $N_{sets}$ is the number of such fully commuting sets and $N_{shots}$ is the number of measurements needed on each circuit to extract the probabilities $P_{|\Psi\rangle}(b_1...b_N)$ with the desired precision.
While the scaling of $N_{sets}$ is addressed in appendix~\ref{App:PauliGrouping}, numerical estimates for the number of measurements required to reach a given precision on the Green’s function are discussed in appendix~\ref{App:ErrorEstimates}.

\subsection{Pauli grouping} \label{App:PauliGrouping}

Let us now estimate $N_{sets}$, i.e. the number of groups of Pauli terms required to evaluate expectation values~\eqref{eq:overlap-H}. 
Terms coming from ${\cal O}^{N + 1}_{ii}$, ${\cal O}^{N - 1}_{ii}$ and one-body terms from ${\cal H}^{N + 1}_{ii}$ and ${\cal H}^{N - 1}_{ii}$ contain only $\mathds{1}$'s and $Z$'s, such that they can all (for all values of $i$) be evaluated together. 
The remaining contributions, coming from $\Gamma_{\!i}$ and $Z_i\Gamma_{\!i}$ terms in Eqs.~\eqref{Pauliterms_H^Np1}-\eqref{Pauliterms_HNm1}, take the forms $P^{(u)}_{kl}$ and $Z_i P^{(u)}_{kl}$, where $P^{(1)}_{kl}, ... , P^{(8)}_{kl}$ are the Pauli terms, acting on qubits $k, \bar{k}, l, \bar{l}$, reading respectively as
\begin{align}
    XXXX, ~~~~ YYYY, ~~~~ XYXY, ~~~~ YXYX, \nonumber \\
    XXYY, ~~~~ YYXX, ~~~~ XYYX, ~~~~ YXXY.
\end{align}

In order to group them, let us define the $8D$ sets $E^{(u)}_i \equiv \{P^{(u)}_{kl}, Z_i P^{(u)}_{kl} \}_{1\leq k<l \leq D,~ k,l\neq i}$. 
These are fully commuting sets, such that each of them requires one basis to evaluate all their terms at once.
For $E^{(1)}_i$, $E^{(2)}_i$, $E^{(3)}_i$, $E^{(4)}_i$, the corresponding measurement basis is composed of $\mathcal{B}_Z \otimes \mathcal{B}_Z$ on qubits $i$ and $\bar{i}$ and respectively $\mathcal{B}_X \otimes \mathcal{B}_X$, $\mathcal{B}_Y \otimes \mathcal{B}_Y$, $\mathcal{B}_X \otimes \mathcal{B}_Y$, $\mathcal{B}_Y \otimes \mathcal{B}_X$ on qubits $k$ and $\bar{k}$ for all $k\neq i$. 
Bases $\mathcal{B}_{\sigma}$ are shown with their corresponding basis rotations in table~\ref{tab:MeasurementBases}.
Sets $E^{(5)}_i$, $E^{(6)}_i$, $E^{(7)}_i$, $E^{(8)}_i$ work differently, since they contain terms having an $X$ and a $Y$ on the same qubit (e.g. in $E^{(5)}_i$, $X_{\!1} X_{\!\bar{1}} Y_{\!2} Y_{\!\bar{2}}$ and $X_{\!2} X_{\!\bar{2}} Y_{\!3} Y_{\!\bar{3}}$). 
The corresponding basis rotations require the use of an extra CNOT gate for each qubit pair $(k,\bar{k})$ with $k\neq i$, as illustrated in table~\ref{tab:MeasurementBases}. 
For example, all terms of the set $E^{(5)}_i$ can be measured in a basis made of $\mathcal{B}_{U}$ in each pair $(k,\bar{k})$ with $k\neq i$.

For this specific system one thus obtains $N_{sets} = 8D + 1$ (i.e., the $8\times D$ sets $E^{(u)}_i$ plus the $1$ set with all terms containing only $\mathds{1}$'s and $Z$'s). 
This can be further reduced by noticing that the reunion of $E^{(1)}_i$, $E^{(2)}_i$, $E^{(5)}_i$, $E^{(6)}_i$ is also fully commuting and its terms can all be measured using $\mathcal{B}_{U}$. Similarly the terms of $E^{(3)}_i$, $E^{(4)}_i$, $E^{(7)}_i$, $E^{(8)}_i$ can all be measured using $\mathcal{B}_{V}$. 
Hence, one can reduce $N_{sets}$ to $2D+1$.
Alternatively, if one wants to avoid the use of extra CNOT gates for measurements, one can divide $E^{(5)}_i$, $E^{(6)}_i$, $E^{(7)}_i$, $E^{(8)}_i$ in sets with e.g. fixed $k$, whose corresponding basis measurement does not require CNOT gates, leading to $N_{sets} = O(D^2)$ to compute the whole Green's function.

\subsection{Computational cost} \label{App:ErrorEstimates}

\begin{figure}
\begin{center}
\includegraphics[width=8.6cm]{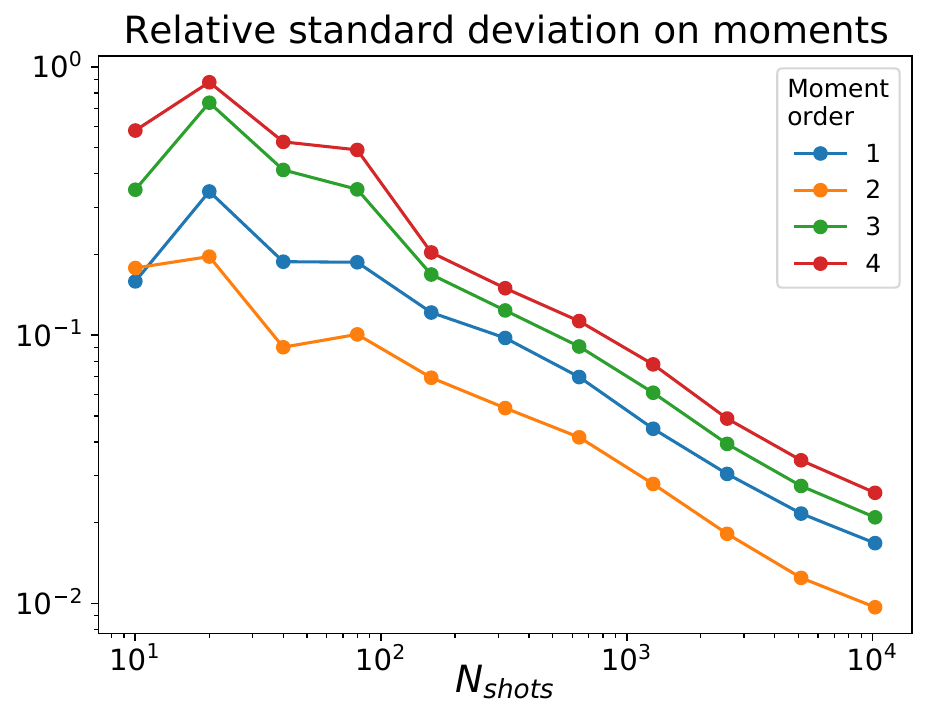}
\end{center}
\centering
\vspace{-0.8 cm}
\caption{Relative standard deviation (standard deviation divided by mean value) on the first moments of the computed Green's function~\eqref{eq:SSD} of a system with $N=8$ particles in $D=8$ doubly-degenerated levels and pairing strength $g=0.6$, computed for several values of $N_{shots}$.
The Pauli terms whose expectation values are required to build the Green's functions are grouped into $N_{sets} = 2D+1$ fully commuting sets as explained in appendix~\ref{App:PauliGrouping}. 
For several values of $N_{shots}$, $N_{shots}$ noiseless quantum measurements are simulated for each of these groups. 
From these $N_{sets} \times N_{shots}$ measurements, the Green's function and the first moments of its spectral strength distribution are computed. 
This procedure is repeated $100$ times in order to estimate the standard deviation and mean value of the resulting moments.}
\label{fig:Nshots_vs_std_moments}
\end{figure}

While the previous sections have addressed the general scaling properties of the expectation values given by~\eqref{eq:overlap-H}, a relevant question concerns the number of measurements that are required in practice to simulate the Green’s function up to a given target error.
Because of the diagonalization step~\eqref{eq:geneigen}, one can not analytically relate a desired precision on the Green's function to errors stemming from finite numbers of $N_{shots}$.
Instead, we estimate such a relation numerically.

To this purpose, we have simulated measurements of the expectation values~\eqref{eq:overlap-H} on a noiseless quantum computer for the case of $N=8$ particles in $D=8$ doubly-degenerated levels and pairing strength $g=0.6$.
For each of the $N_{sets}$ groups, we have performed experiments with different numbers of measurements $N_{shots}$ and subsequently reconstructed the corresponding Green's function.
For each value of $N_{shots}$, the standard deviation with respect to the exact solution for the first four moments of the spectral strength distribution has been determined by repeating the experiment 100 times\footnote{
%To avoid pathologically small overlaps originating from numerical noise, we have excluded samples with overlaps smaller than $(2 N_{shots})^{-1}$.
In some experiments, the error on the overlaps~(\ref{eq:overlap}) can be such that the overlap matrices are not positive definite and thus the generalized eigenvalue problem~(\ref{eq:geneigen}) cannot be solved. These pathological cases have been excluded from our analysis.}.

Fig.~\ref{fig:Nshots_vs_std_moments} displays the relative standard deviation of the four moments as a function of the number of measurements $N_{shots}$.
One observes that the error decreases as $1/\sqrt{N_{shots}}$ for all curves, reaching values around $10\%$ for $N_{shots} \approx 200$.
In order to reach errors of the order of $1\%$, one needs instead to increase the number of measurements to $N_{shots} \approx 10^4$.

We have performed tests with different numbers of particles and levels, which showed a similar behavior. 
One can thus assume the dependence on $N_{shots}$ emerging from Fig.~\ref{fig:Nshots_vs_std_moments} to be independent of the system size $D$.
As discussed above, the total number of required measurements will be dictated by the product $N_{sets} \times N_{shots}$, with $N_{sets}$ driving the scaling as a function of $D$.

\color{black}

\end{document}